\documentclass[12pt]{article}
\usepackage{pdproc,epsfig}
\def\numu{\nu_{\mu}}
\def\numubar{\bar\nu_{\mu}}
\def\nue{\nu_e}
\def\nuebar{\bar\nu_e}

\def\gtwid{\mathrel{\raise.3ex\hbox{$>$\kern-.75em\lower1ex\hbox{$\sim$}}}}
\def\ltwid{\mathrel{\raise.3ex\hbox{$<$\kern-.75em\lower1ex\hbox{$\sim$}}}}

\newcommand{\text}[1]{\mbox{#1}}

\begin{document}

% \draft command makes pacs numbers print
%\draft
 
%\begin{titlepage}
 
\centerline{\Large A Letter of Intent to Build a MiniBooNE Near Detector:BooNE}
 
\vspace{0.2in}
 
%\date{\today}

\centerline{October 12, 2009} 

\bigskip

\centerline{I. Stancu}

\centerline{\it University of Alabama, Tuscaloosa, AL 35487}

\bigskip

\centerline{Z. Djurcic}

\centerline{\it Argonne National Laboratory, Argonne, IL 60439}

\bigskip

\centerline{D. Smith}

\centerline{\it Embry-Riddle Aeronautical University, Prescott, AZ 86301}

\bigskip

\centerline{R. Ford, T. Kobilarcik, W. Marsh, \& C. D. Moore}

\centerline{\it Fermi National Accelerator Laboratory, Batavia, IL 60510}

\bigskip

\centerline{J. Grange, B. Osmanov, \& H. Ray}

\centerline{\it University of Florida, Gainesville, FL 32611}

\bigskip

\centerline{G. T. Garvey, J. A. Green, W. C. Louis, C. Mauger, G. B. Mills, Z. Pavlovic,} 
\centerline{R. Van de Water, D. H. White, \& G. P. Zeller}
                                                                                       
\centerline{\it Los Alamos National Laboratory, Los Alamos, NM 87545}                                                                                       
\bigskip

\centerline{W. Metcalf}

\centerline{\it Louisiana State University, Baton Rouge, LA 70803}

\bigskip

\centerline{B. P. Roe}

\centerline{\it University of Michigan, Ann Arbor, MI 48109}

\bigskip

\centerline{A. A. Aguilar-Arevalo}

\centerline{\it Instituto de Ciencias Nucleares, Universidad Nacional Autónoma de M\'exico, M\'exico D.F. M\'exico}

\bigskip
 
\section{ \bf Executive Summary}

There is accumulating evidence for a difference between neutrino and
antineutrino oscillations at the $\sim 1$ eV$^2$ scale. 
The MiniBooNE experiment observes an unexplained excess of electron-like
events at low energies in neutrino mode \cite{mb_osc},
which may be due, for example, to either a neutral
current radiative interaction \cite{hhh}, sterile neutrino decay \cite{gninenko}, or 
to neutrino oscillations involving sterile neutrinos
\cite{sorel,weiler,goldman,maltoni,nelson,paes} 
and which may be related to the LSND signal \cite{lsnd}. No excess 
of electron-like events ($-0.5 \pm 7.8 \pm 8.7$), 
however, is observed so far
at low energies in antineutrino mode \cite{mb_anti}. Furthermore, global
3+1 and 3+2 sterile neutrino
fits to the world neutrino and antineutrino data \cite{georgia} suggest
a difference between neutrinos and antineutrinos with
significant ($\sin^22\theta_{\mu \mu} \sim 35\%$) $\bar \nu_\mu$ disappearance.
In order to test whether the low-energy excess is due to neutrino 
oscillations and whether there is a difference between $\nu_\mu$ and $\bar \nu_\mu$
disappearance, we propose building a second MiniBooNE detector at (or moving
the existing MiniBooNE detector to) a
distance of $\sim 200$ m from the Booster Neutrino Beam (BNB) production
target. With identical detectors at different distances, most of the
systematic errors will cancel when taking a ratio of events in the
two detectors, as the neutrino flux varies as $1/r^2$ to a calculable approximation.
This will allow sensitive tests of oscillations for both $\nu_e$ and
$\bar \nu_e$ appearance and $\nu_\mu$ and $\bar \nu_\mu$
disappearance. Furthermore, a comparison between oscillations
in neutrino mode and antineutrino mode will allow a sensitive search
for CP and CPT violation in the lepton sector at short baseline
($\Delta m^2 > 0.1$ eV$^2$). Finally, by
comparing the rates for a neutral current (NC) reaction, such as NC
$\pi^0$ scattering or NC elastic scattering,
a direct search for sterile neutrinos will be made.
The initial amount of running time requested for the near detector 
will be a total of $\sim 2$E20 POT divided between
neutrino mode and antineutrino mode, which will provide statistics 
comparable to what has already been collected in the far detector.
A thorough understanding of this short-baseline physics will be of
great importance to future long-baseline oscillation experiments.

\section{Introduction}
Evidence for neutrino oscillations comes from solar-neutrino
\cite{homestake98,sage99,gallex99,sk02,sno} and
reactor-antineutrino experiments \cite{kamland}, which have
observed $\nu_e$ disappearance at $\Delta m^2 \sim 8\times 10^{-5}$
eV$^2$, and atmospheric-neutrino
\cite{kam,sk98,soudan99,macro01} and long-baseline
accelerator-neutrino
experiments \cite{k2k03,minos06}, which have observed $\nu_\mu$ disappearance
at $\Delta m^2 \sim 3\times 10^{-3}$ eV$^2$.
In addition, the LSND experiment \cite{lsnd} has presented
evidence for $\bar \nu_\mu \rightarrow \bar
\nu_e$ oscillations at the $\Delta m^2 \sim 1$ eV$^2$ scale.
If all three phenomena are caused by neutrino oscillations, these
three $\Delta m^2$ scales cannot be accommodated within an extension of
the Standard Model with only three neutrino mass eigenstates.
An explanation of all three mass scales with neutrino oscillations
requires the addition of one or more sterile neutrinos
\cite{sorel,weiler,goldman,maltoni,nelson,paes} or
further extensions of the Standard Model ({\it e.g.,} \cite{katori}).

The MiniBooNE experiment was designed to test the
neutrino oscillation
interpretation of the LSND signal in both neutrino and
antineutrino modes.
%Fig. \ref{mb_photo} shows a photograph of the inside
%of the MiniBooNE detector before the tank was sealed and filled with
%mineral oil.
MiniBooNE has approximately the same $L/E_\nu$ as
LSND but with an order of magnitude higher baseline and energy. Due to
the higher energy and dissimilar event signature,
MiniBooNE systematic errors are completely different from LSND errors.
MiniBooNE's updated
oscillation results in neutrino mode \cite{mb_osc} show no significant
excess of events at higher energies; however, a sizeable excess of events
is observed at lower energies. Although the
excess energy shape does not fit simple two-neutrino oscillations, the
number of excess events agrees approximately with the LSND expectation.
At present, with 3.4E20 POT in antineutrino
mode, MiniBooNE observes no excess so far
at lower energies, while at higher energies the data are inconclusive with respect
to antineutrino oscillations suggested by the LSND data \cite{lsnd}.

\begin{figure}
\centerline{\includegraphics[height=7in]{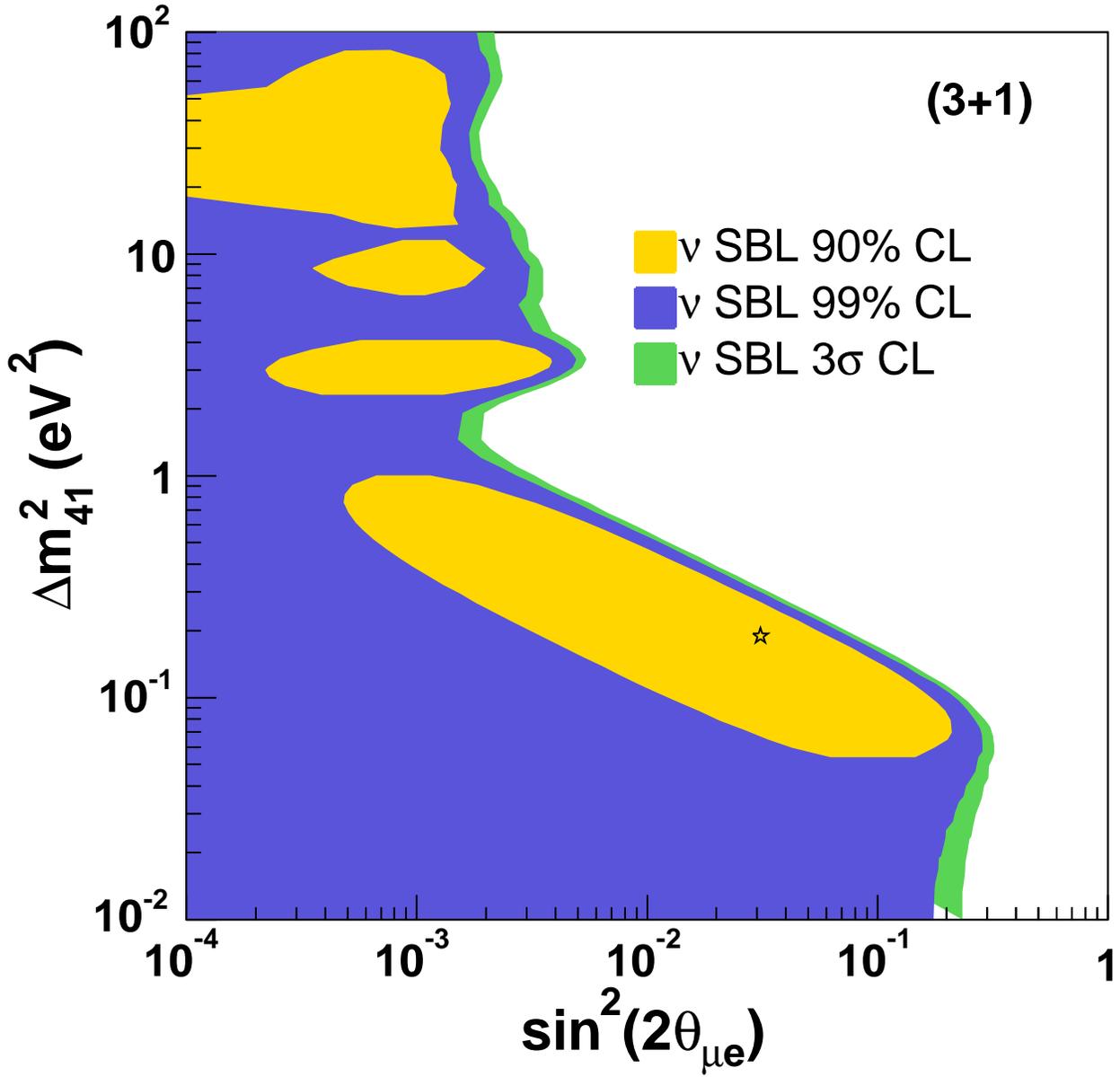}}
\caption{\label{global_fit_nu_app} \em The allowed region for $\nu_e$
appearance from a global fit to the world neutrino data, assuming
a 3+1 sterile neutrino model. The star indicates the best-fit point at $\Delta m^2_{41} = 0.19$ eV$^2$ and 
$\sin^22\theta_{\mu e} = 0.031$.}
\end{figure}

\begin{figure}
\centerline{\includegraphics[height=7in]{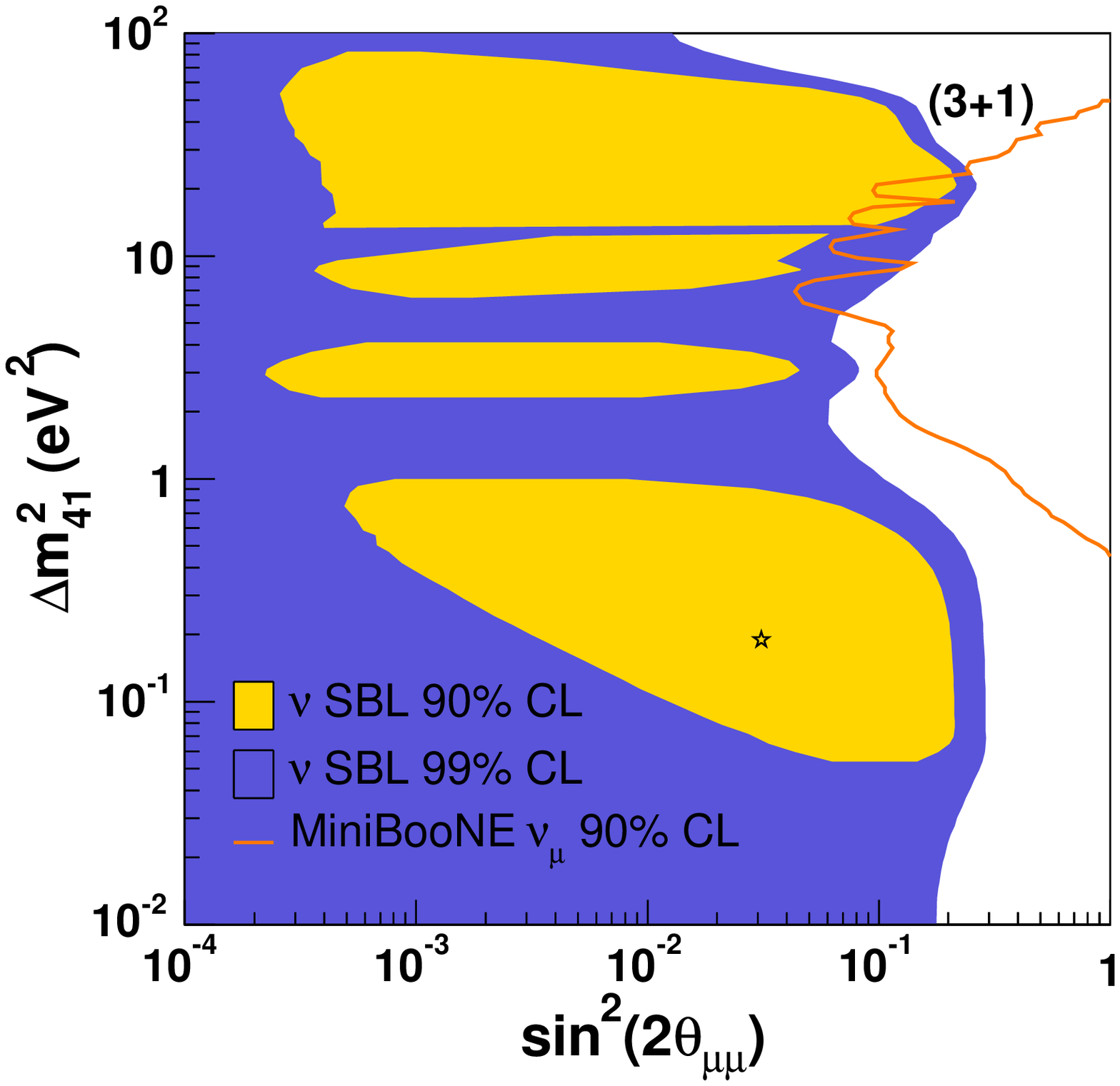}}
\caption{\label{global_fit_nu_dis} \em The allowed region for $\nu_\mu$
disappearance from a global fit to the world neutrino data, assuming
a 3+1 sterile neutrino model. The star indicates the best-fit point at $\Delta m^2_{41} = 0.19$ eV$^2$ and 
$\sin^22\theta_{\mu \mu} = 0.031$.}
\end{figure}

\begin{figure}
\centerline{\includegraphics[height=7in]{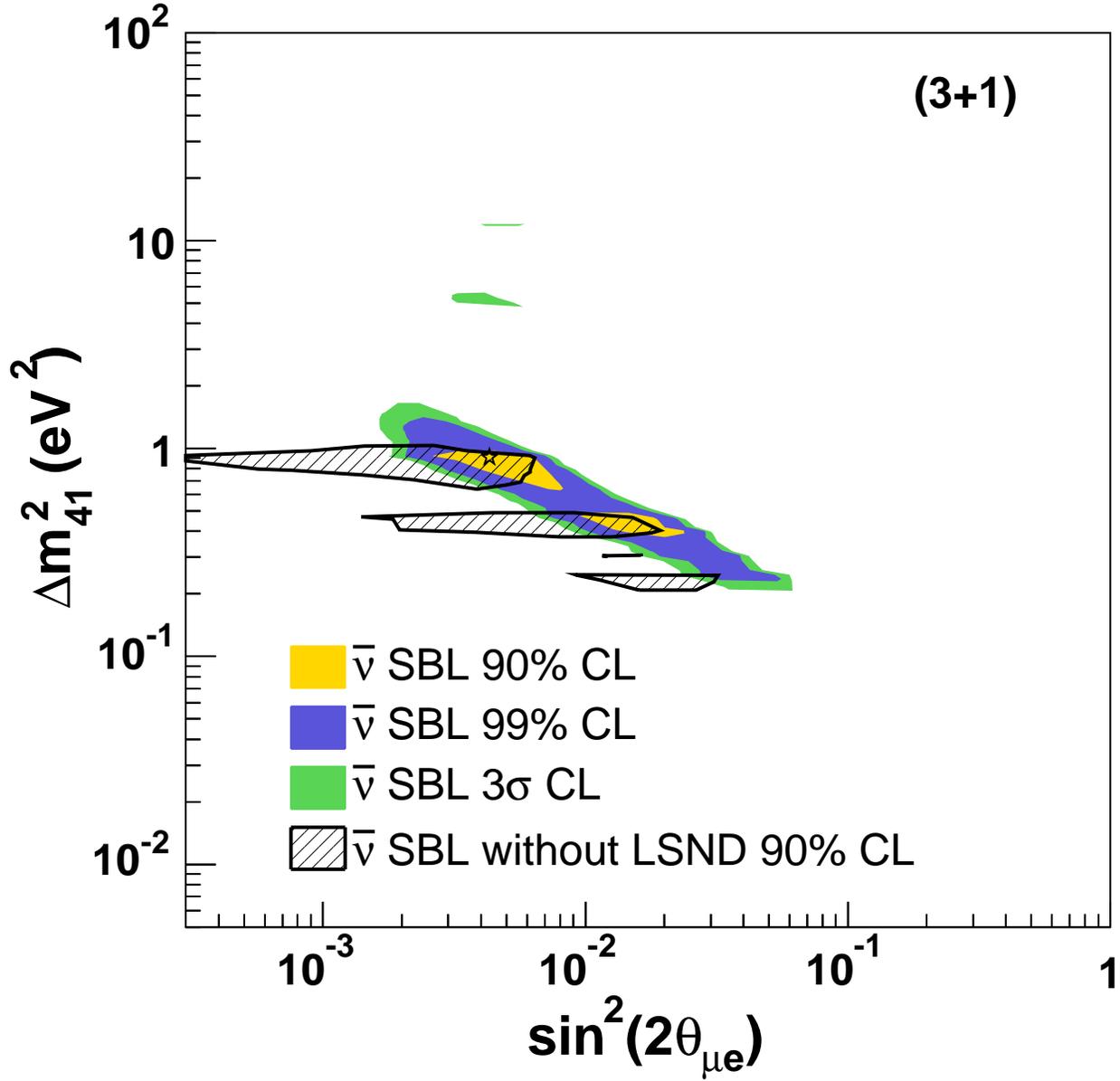}}
\caption{\label{global_fit_antinu_app} \em The allowed region for $\bar \nu_e$
appearance from a global 
fit to the world antineutrino data, assuming
a 3+1 sterile antineutrino model. The star indicates the best-fit point at $\Delta m^2_{41} = 0.915$ eV$^2$ and $\sin^22\theta_{\mu e} = 0.0043$. Also shown
is the global fit to the world antineutrino data without LSND.}
\end{figure}

\begin{figure}
\centerline{\includegraphics[height=7in]{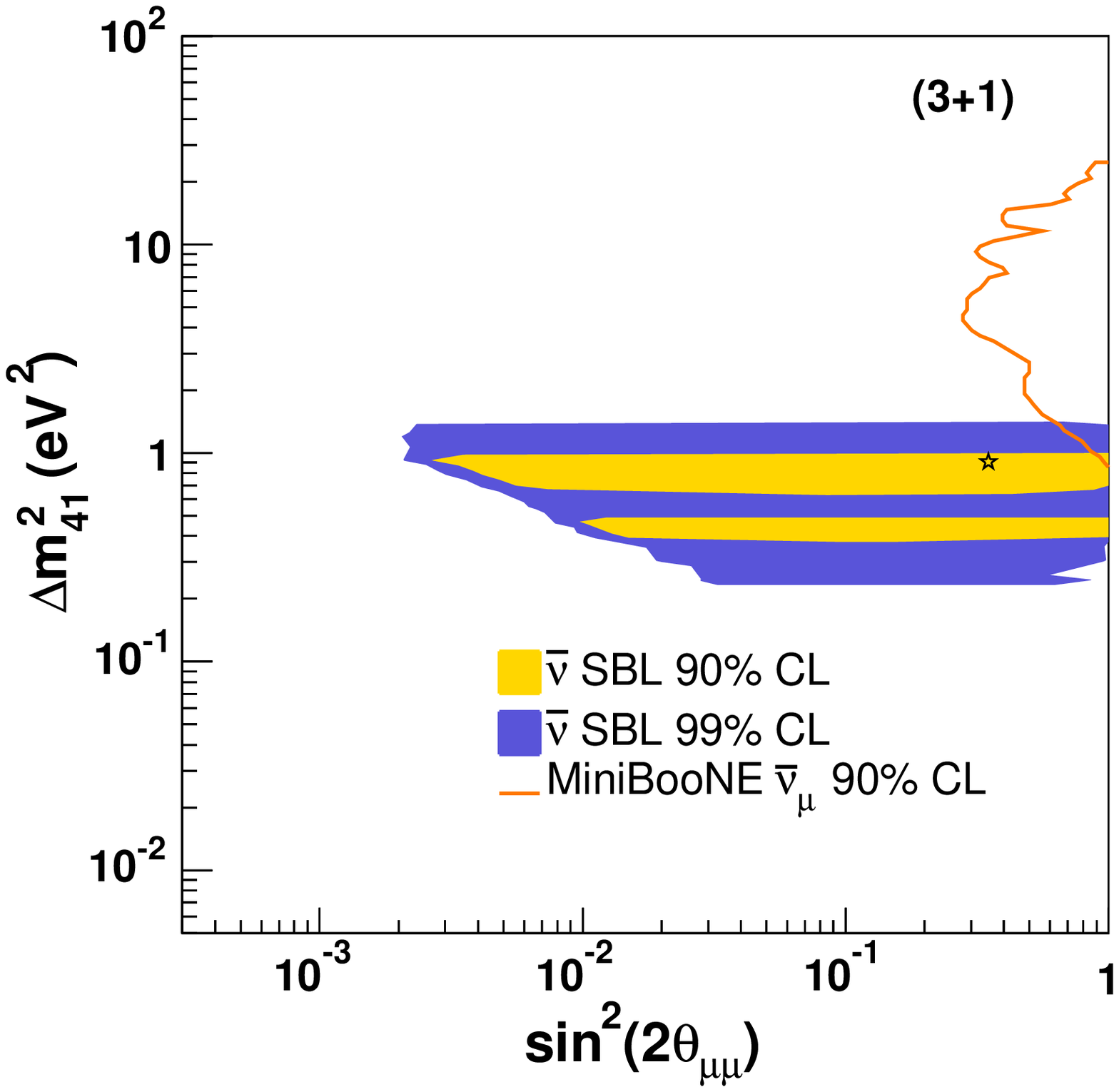}}
\caption{\label{global_fit_antinu_dis} \em The allowed region for $\bar \nu_\mu$
disappearance from a global
fit to the world antineutrino data, assuming
a 3+1 sterile antineutrino model. The star indicates the best-fit point at 
$\Delta m^2_{41} = 0.915$ eV$^2$ and $\sin^22\theta_{\mu \mu} = 0.35$.}
\end{figure}

A global fit to the world neutrino oscillation data has been performed in terms
of 3+1 and 3+2 sterile neutrino models \cite{georgia}. Although a large 
incompatibility is observed when fitting all of the data, there is much less
tension when fitting the world neutrino and antineutrino data separately. 
Figs. \ref{global_fit_nu_app}, \ref{global_fit_nu_dis},
\ref{global_fit_antinu_app}, and \ref{global_fit_antinu_dis}
show the allowed regions from global fits to the world neutrino and antineutrino data, assuming
3+1 sterile neutrino models. The
best neutrino fit occurs at $\Delta m^2_{41} = 0.19$ eV$^2$ with $\sin^22\theta_{\mu e} = 0.031$,
$\sin^22\theta_{\mu\mu} = 0.031$, and $\sin^22\theta_{ee}=0.034$, corresponding to 
a $\chi^2 = 90.5/90$ DF (probability = 47\%). The
best antineutrino fit occurs at $\Delta m^2_{41} = 0.915$ eV$^2$ with $\sin^22\theta_{\mu e} = 0.0043$,
$\sin^22\theta_{\mu\mu} = 0.35$, and $\sin^22\theta_{ee}=0.043$, corresponding to 
a $\chi^2 = 87.9/103$ DF (probability = 86\%). The antineutrino fit is
dominated by the LSND data \cite{lsnd}; however, 
Fig. \ref{global_fit_antinu_app} also
shows that the global antineutrino data without LSND is consistent with the
LSND signal and has a closed 90\% CL contour around the LSND best-fit point. 
CPT-conserving oscillation scenarios appear insufficient to explain all of 
the data. As stated in reference \cite{georgia},
``CPT-violating oscillations or effective CPT violation \cite{gabriela,paes}
may succeed in reconciling
all short-baseline oscillation signatures, and should be pursued''. 

%%\begin{figure}[ht]
%\begin{minipage}{14pc}
%%\centerline{\includegraphics[height=4.0in]{boone_photo.eps}}
%\centerline{\psfig{file=boone_photo.eps,width=14.cm}}
%\includegraphics[width=38pc]{boone_photo.eps}
%%\caption{\label{mb_photo} A photograph of the inside
%%of the MiniBooNE detector before the tank was sealed and filled with
%%mineral oil.}
%\end{minipage}\hspace{2pc}%
%\begin{minipage}{14pc}
%\includegraphics[width=14pc]{name.eps}
%\caption{\label{label}Figure caption for second of two sided figures.}
%\end{minipage}
%%\end{figure}

\section{MiniBooNE}

\subsection{Description of the Experiment}

\begin{figure}
%\begin{minipage}{14pc}
\centerline{\includegraphics[height=1.25in]{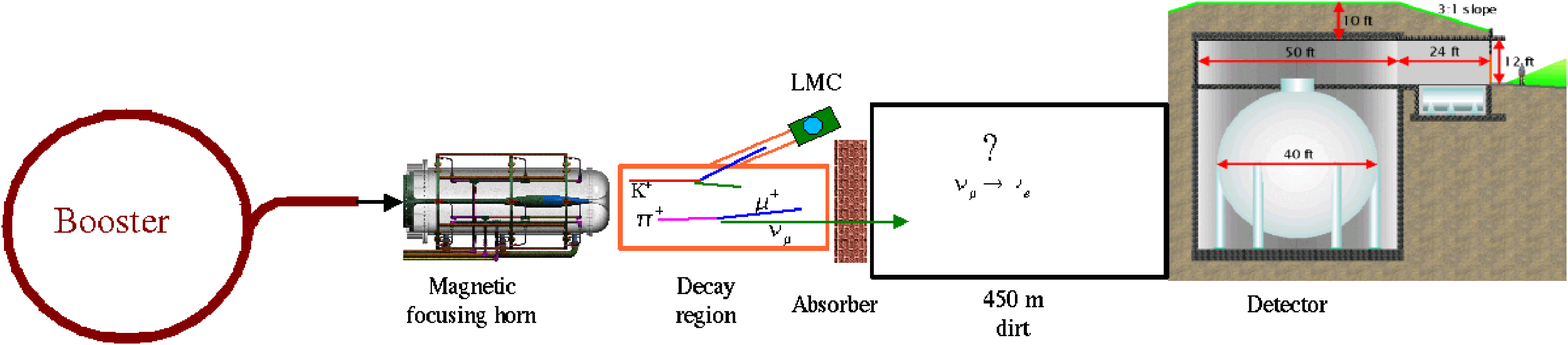}}
%\centerline{\psfig{file=geography.eps,width=18.cm}}
%\includegraphics[width=38pc]{geography.eps}
\caption{\label{schematic} \em A schematic drawing of the MiniBooNE experiment.}%\end{minipage}\hspace{2pc}%
%\begin{minipage}{14pc}
%\includegraphics[width=14pc]{name.eps}
%\caption{\label{label}Figure caption for second of two sided figures.}
%\end{minipage}
\end{figure}

A schematic drawing
of the MiniBooNE experiment at FNAL is shown in Fig. \ref{schematic}.
The experiment is fed by 8-GeV kinetic energy
protons from the Booster that interact in
a 71-cm long Be target located at the upstream end of a magnetic
focusing horn. The horn pulses with a current of 174 kA and, depending on
the polarity, either focuses $\pi^+$ and $K^+$ and defocuses $\pi^-$ and $K^-$
to form a pure neutrino beam or focuses $\pi^-$ and $K^-$ and
defocuses $\pi^+$ and $K^+$ to form a somewhat pure antineutrino beam.
The produced pions and kaons decay in a 50-m long pipe, and a fraction 
of the neutrinos and antineutrinos \cite{mb_flux}
interact in the MiniBooNE
detector, which is located 541 m downstream of the Be target. For the MiniBooNE
results presented here, a total of $6.5 \times 10^{20}$
POT were collected in neutrino mode and $3.4 \times 10^{20}$
POT were collected in antineutrino mode.

\begin{figure}
%\begin{minipage}{14pc}
\centerline{\includegraphics[height=3.5in]{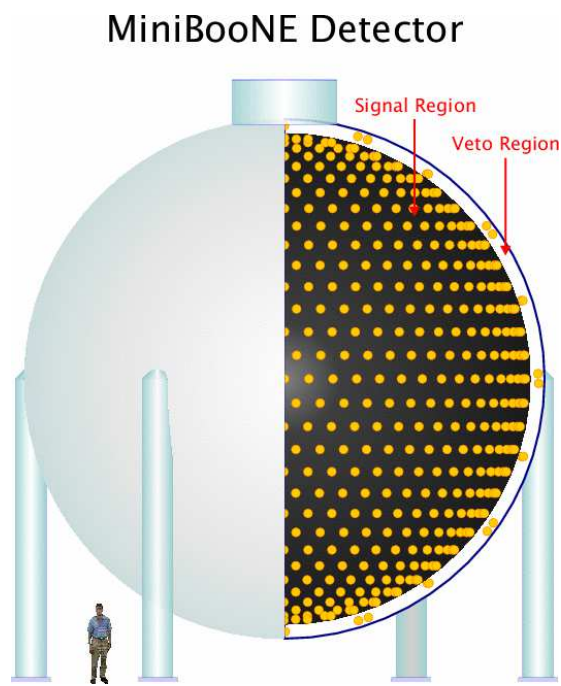}}
%\centerline{\psfig{file=boone_tank_new.epsi,width=12.cm}}
%\includegraphics[width=20pc]{boone_tank_new.epsi}
\caption{\label{mb_schematic} \em A schematic drawing of the MiniBooNE detector.}
%\end{minipage}\hspace{2pc}%
%\begin{minipage}{14pc}
%\includegraphics[width=14pc]{name.eps}
%\caption{\label{label}Figure caption for second of two sided figures.}
%\end{minipage}
\end{figure}

The MiniBooNE detector \cite{mb_detector} consists of a 12.2-m diameter
spherical tank filled with approximately 800 tons of mineral oil ($CH_2$).
A schematic drawing
of the MiniBooNE detector is shown in Fig. \ref{mb_schematic}.
There are
a total of 1280 8-inch detector phototubes (covering 10\% of the surface area)
and 240 veto phototubes. The fiducial volume has a 5-m radius and
corresponds to approximately 450 tons. Only $\sim 2\%$ of the phototube
channels failed over the course of the run.

\subsection{MiniBooNE Cross Section Results}

\begin{figure}
%\begin{minipage}{18pc}
\centerline{\includegraphics[height=3.5in]{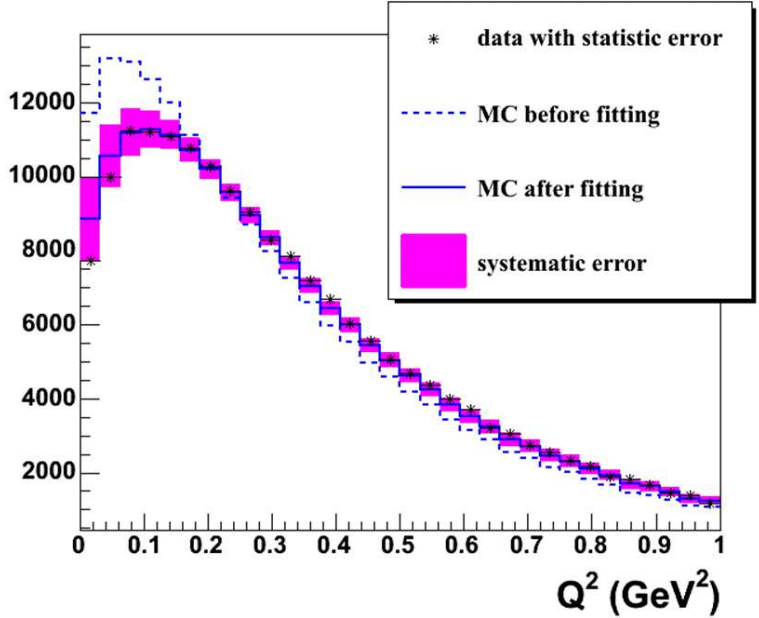}}
%\centerline{\psfig{file=ccqe.eps,width=14.cm}}
%\includegraphics[width=38pc]{ccqe.eps}
\caption{\label{ccqe} \em The $\nu_\mu$ CCQE $Q^2$ distribution for
data (points with error bars) compared to the MC
simulation (histograms).}
\end{figure}

MiniBooNE has published two cross section results. First,
MiniBooNE has made a high-statistics measurement of $\nu_\mu$ charged-current
quasi-elastic (CCQE) scattering events \cite{mb_ccqe}.
Fig. \ref{ccqe} shows the $\nu_\mu$ CCQE $Q^2$ distribution for
data (points with error bars) compared to a MC
simulation (histograms). A strong disagreement between the data and the
original simulation (dashed histogram) was first observed. However, by
increasing the axial mass, $M_A$,
to $1.23 \pm 0.20$ GeV and by introducing a new
variable, $\kappa=1.019 \pm 0.011$, where $\kappa$ is the increase in the
incident proton threshold, the agreement between data and the
simulation (solid histogram) is greatly improved. It is impressive that
such good agreement is obtained by adjusting these two variables.

\begin{figure}
\centerline{\includegraphics[height=3.in]{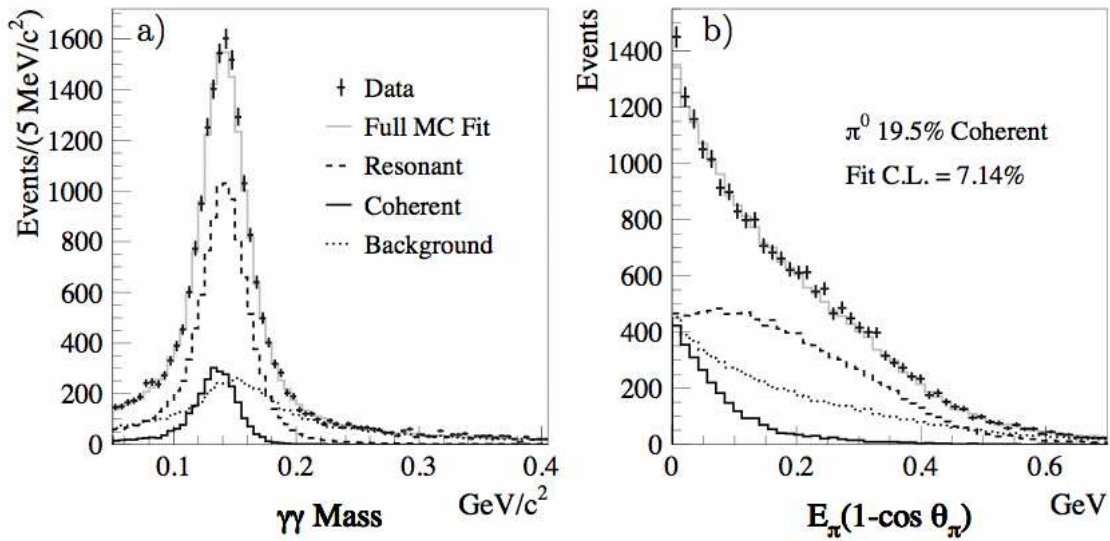}}
%\centerline{\psfig{file=ncpi0.eps,width=14.cm}}
%\includegraphics[width=38pc]{ncpi0.eps}
\caption{\label{ncpi0} \em The neutral-current $\pi^0$ $\gamma \gamma$
mass and $E_\pi(1-\cos \theta_\pi)$ distributions for data
(points with error bars) compared to the MC
simulation (histograms).}
\end{figure}

MiniBooNE has also collected the world's largest sample of neutral-current
$\pi^0$ events \cite{mb_ncpi0},
as shown in Fig. \ref{ncpi0}. By fitting the $\gamma \gamma$
mass and $E_\pi(1-\cos \theta_\pi)$ distributions, the fraction of
$\pi^0$ produced coherently is determined to be $19.5 \pm 1.1 \pm 2.5\%$.
Excellent agreement is obtained between data and MC
simulation.

\subsection{Neutrino Oscillation Event Selection}

MiniBooNE searches for $\nu_\mu \rightarrow \nu_e$ oscillations by
measuring the rate of $\nu_e C \rightarrow e^- X$ CCQE events and
testing whether the measured rate is consistent with the estimated
background rate.
To select candidate $\nu_e$ CCQE events, an initial selection is
first applied:
$>200$ tank hits, $<6$ veto hits, reconstructed time within the
neutrino beam spill, reconstructed vertex radius $<500$ cm,
and visible energy $E_{vis}>140$ MeV.
It is then required
that the event vertex reconstructed assuming an outgoing electron and the track
endpoint reconstructed assuming an outgoing muon
occur at radii $<500$ cm and $<488$ cm, respectively, to
ensure good event reconstruction and efficiency for possible muon decay
electrons. Particle identification (PID) cuts are then applied
to reject muon and $\pi^0$ events.
Several improvements have been made to the
neutrino oscillation data
analysis since the initial data was published \cite{mb_osc},
including an improved background estimate, an additional
fiducial volume cut that greatly reduces the background from events produced
outside the tank (dirt events), and an increase in the
data sample from $5.579 \times 10^{20}$ POT to $6.462 \times 10^{20}$ POT.
A total of 89,200 neutrino
events pass the initial selection,
while 1069 events pass the complete event selection of the final analysis
with $E_\nu^{QE}>200$ MeV, where $E_\nu^{QE}$ is the reconstructed neutrino
energy.

\subsection{Neutrino Oscillation Signal and Background Reactions}

Table \ref{signal_bkgd} shows the
expected number of candidate $\nu_e$ CCQE background events with
$E_\nu^{QE}$ between $200 - 300$ MeV, $300 - 475$ MeV, and
$475 - 1250$ MeV after
the complete event selection of the final analysis.
The background estimate includes
antineutrino events, representing $<2\%$ of the total.
The total expected backgrounds for the three energy regions are
$186.8 \pm 26.0$ events, $228.3 \pm 24.5$ events,
and $385.9 \pm 35.7$ events, respectively. For $\nu_\mu \rightarrow
\nu_e$ oscillations at the best-fit LSND solution of $\Delta m^2 =1.2$ eV$^2$
and $\sin^22\theta = 0.003$, the expected number of $\nu_e$ CCQE signal
events for
the three energy regions are 7 events, 37 events, and 135 events, respectively.

\begin{table}
%\begin{table}[h]
\begin{center}
\caption{\label{signal_bkgd} \em The expected number of events
in the $200<E_\nu^{QE}<300$ MeV, $300<E_\nu^{QE}<475$ MeV,
and $475<E_\nu^{QE}<1250$ MeV
energy ranges from all of the significant backgrounds
after the complete event selection of the final analysis.
Also shown are the expected number of $\nu_e$ CCQE signal events for
two-neutrino oscillations at the LSND best-fit solution.}
%\begin{ruledtabular}
\vspace{0.1in}
\begin{tabular}{|c|c|c|c|}
\hline
Process&$200-300$&$300-475$&$475-1250$ \\
\hline
%Beam Unrelated&2 \\
%\hline
$\nu_\mu$ CCQE&9.0&17.4&11.7 \\
$\nu_\mu e \rightarrow \nu_\mu e$&6.1&4.3&6.4 \\
NC $\pi^0$&103.5&77.8&71.2 \\
NC $\Delta \rightarrow N \gamma$&19.5&47.5&19.4 \\
Dirt Events&11.5&12.3&11.5 \\
Other Events&18.4&7.3&16.8 \\
\hline
$\nu_e$ from $\mu$ Decay&13.6&44.5&153.5 \\
$\nu_e$ from $K^+$ Decay&3.6&13.8&81.9 \\
$\nu_e$ from $K^0_L$ Decay&1.6&3.4&13.5 \\
\hline
Total Background &$186.8 \pm 26.0$&$228.3 \pm 24.5$&$385.9 \pm 35.7$ \\
\hline
LSND Best-Fit Solution&$7 \pm 1$&$37 \pm 4$&$135 \pm 12$ \\
\hline
\end{tabular}
\end{center}
%\vspace{-0.2in}
%\end{ruledtabular}
\end{table}

\subsection{Updated Neutrino Oscillation Results}

\begin{figure}
\centerline{\includegraphics[height=3.in]{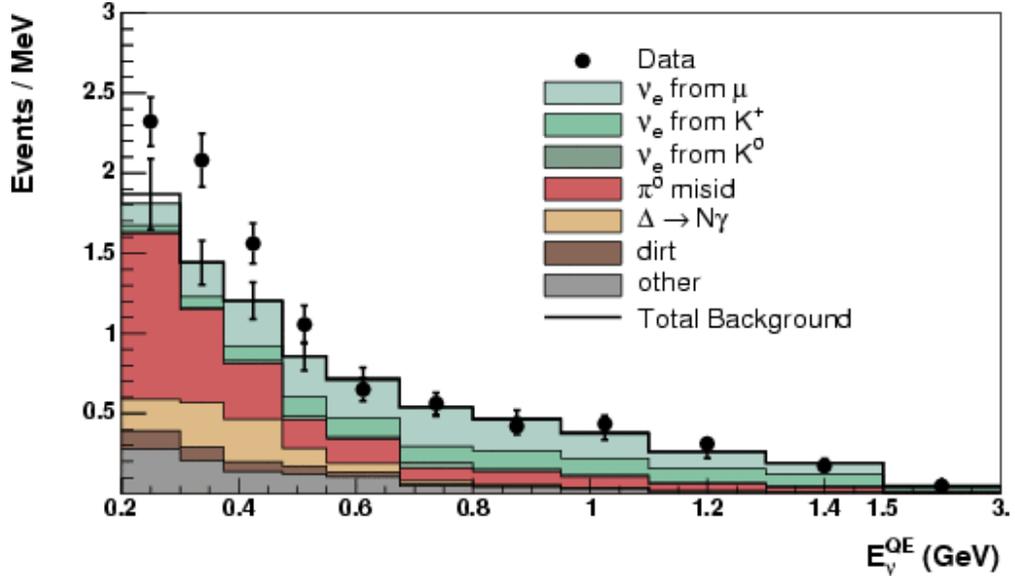}}
%\centerline{\psfig{file=figure1_enuqe.eps,width=14.cm}}
%\centerline{\psfig{file=fulldata_stacked_plot_200fit_dirtcut.eps,width=14.cm}}
\caption{\label{osc} \em The MiniBooNE reconstructed neutrino energy distribution
for candidate $\nu_e$ data events (points with error bars) compared to
the Monte Carlo simulation (histogram).}
% \cite{mb_osc}.}
%\end{minipage}\hspace{2pc}%
%\begin{minipage}{14pc}
%\includegraphics[width=14pc]{name.eps}
%\caption{\label{label}Figure caption for second of two sided figures.}
%\end{minipage}
\end{figure}

\begin{figure}
\centerline{\includegraphics[height=3.in]{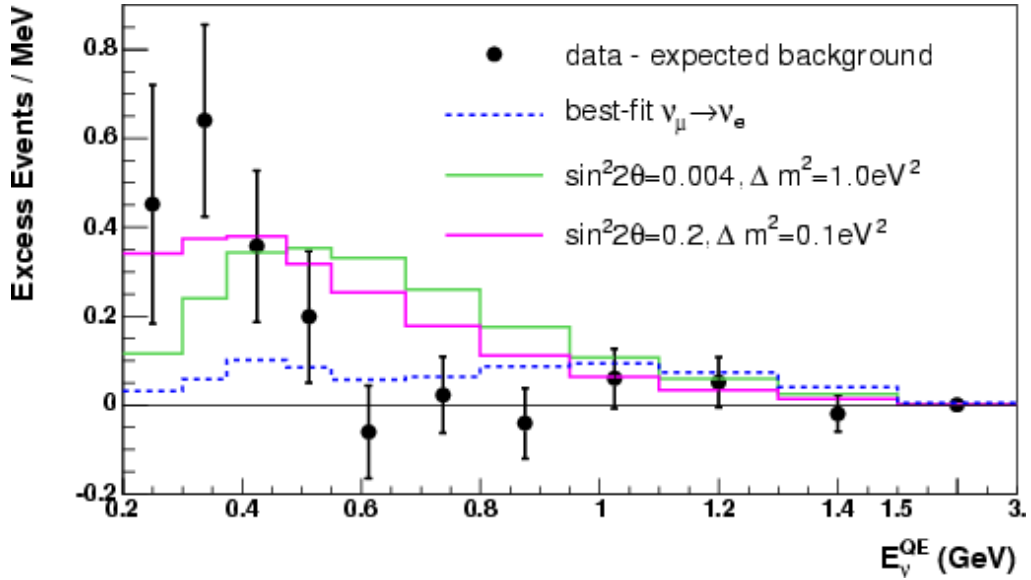}}
%\centerline{\psfig{file=figure2_enuqe.eps,width=16.cm}}
%\includegraphics[width=38pc]{fulldata_stacked_plot_200fit_dirtcut_excess.eps}
\caption{\label{osc_excess} \em The event excess as a function of $E_\nu^{QE}$.
Also shown are
the expectations from
the best oscillation fit ($\sin^22\theta = 0.0017$, $\Delta m^2 = 3.14$ eV$^2$)
and from
neutrino oscillation parameters in the LSND
allowed region. The error bars include both statistical and systematic errors.}
\end{figure}

Fig. \ref{osc} shows the reconstructed neutrino energy distribution
for candidate $\nu_e$ data events (points with error bars) compared to
the MC simulation (histogram) \cite{mb_osc}, while
Fig. \ref{osc_excess} shows the event excess as a function of reconstructed
neutrino energy. Good agreement
between the data and the MC simulation is obtained for
$E_\nu^{QE} > 475$ MeV; however, an unexplained excess of electron-like
events is observed for $E_\nu^{QE} < 475$ MeV. As shown in Fig. \ref{osc_excess},
the magnitude of the excess is very similar to what is expected from
neutrino oscillations based on the LSND signal. Although the shape of the
excess is not consistent with simple two-neutrino oscillations, more
complicated oscillation models \cite{sorel,weiler,goldman,maltoni,nelson,paes}
or sterile neutrino decay \cite{gninenko}
have shapes that may be consistent with the LSND signal. A test of the sterile
neutrino decay model \cite{gninenko} can be performed by searching for the
decay $D^+_s \rightarrow \mu^+ \nu_h$, where the heavy sterile neutrino $\nu_h$
has a mass around 500 MeV.

\begin{figure}
%\centerline{\includegraphics[height=3.2in]{evis_excess.eps}}
\centerline{\includegraphics[height=6.in,angle=-90]{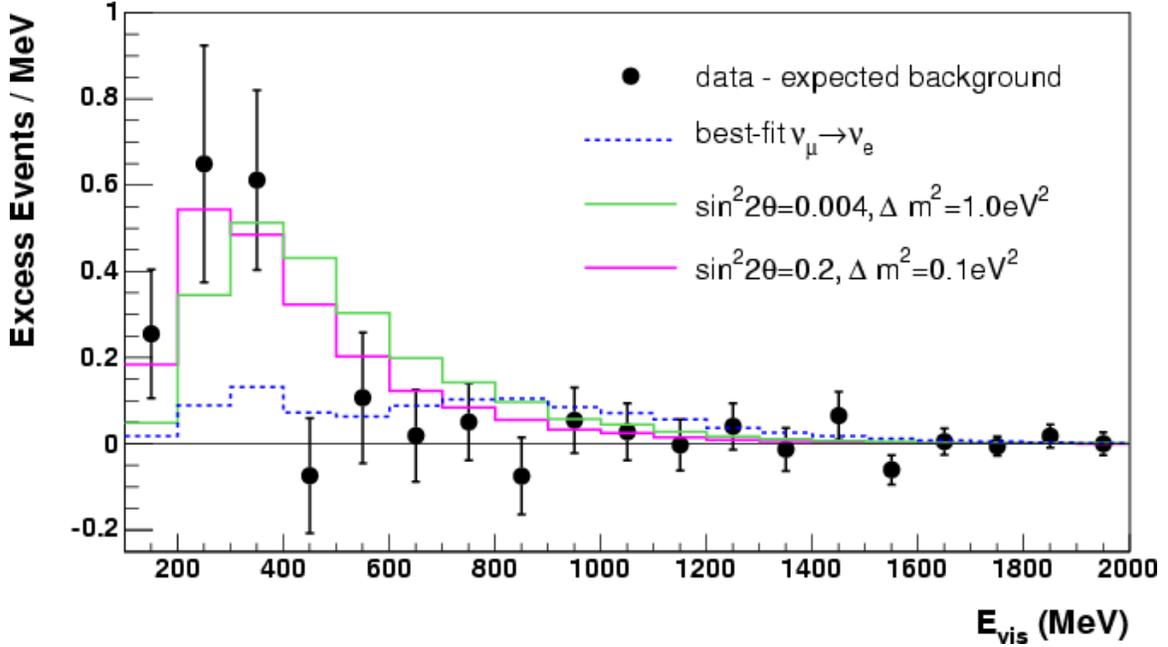}}
\caption{\em The event neutrino excess as a
function of $E_{vis}$ for $E_\nu^{QE} > 200$ MeV.
Also shown are
the expectations from
the best oscillation fit ($\sin^22\theta = 0.0017$, $\Delta m^2 = 3.14$ eV$^2$)
and from
neutrino oscillation parameters in the LSND
allowed region. The error bars include both statistical and systematic errors.}
\label{data_mc3}
\end{figure}

Table \ref{mb_events} shows the number of data,
background, and excess events for different
$E_\nu^{QE}$ ranges, together with the
excess significance.
For the final analysis, an excess of $128.8 \pm 20.4 \pm 38.3$ events
is observed for $200<E_\nu^{QE}<475$ MeV. For the
entire $200<E_\nu^{QE}<1250$ MeV energy region, the excess is
$151.0 \pm 28.3 \pm 50.7$ events.
As shown in Fig. \ref{data_mc3}, the event excess
occurs for $E_{vis} <400$ MeV, where $E_{vis}$ is the visible energy.

\begin{table}
%\begin{table}[h]
%\begin{center}
\centering
\caption{\label{mb_events} \em The number of data,
background, and excess events for different
$E_\nu^{QE}$ ranges, together with the
significance of the excesses in neutrino mode.}
%%\begin{ruledtabular}
\vspace{0.1in}
\begin{tabular}{|c|c|}
\hline
Event Sample&Final Analysis \\
\hline
$200-300$ MeV& \\
Data&232 \\
Background&$186.8 \pm 13.7 \pm 22.1$ \\
Excess&$45.2 \pm 13.7 \pm 22.1$ \\
Significance&$1.7 \sigma$ \\
\hline
$300-475$ MeV& \\
Data&312 \\
Background&$228.3 \pm 15.1 \pm 19.3$ \\
Excess&$83.7 \pm 15.1 \pm 19.3$ \\
Significance&$3.4 \sigma$ \\
\hline
$200-475$ MeV& \\
Data&544 \\
Background&$415.2 \pm 20.4 \pm 38.3$ \\
Excess&$128.8 \pm 20.4 \pm 38.3$ \\
Significance&$3.0 \sigma$ \\
\hline
$475-1250$ MeV& \\
Data&408 \\
Background&$385.9 \pm 19.6 \pm 29.8$ \\
Excess&$22.1 \pm 19.6 \pm 29.8$ \\
Significance&$0.6 \sigma$ \\
\hline
\end{tabular}
%\end{center}
%%\end{ruledtabular}
%\end{center}
\end{table}

\begin{figure}
\centerline{\includegraphics[height=3.in]{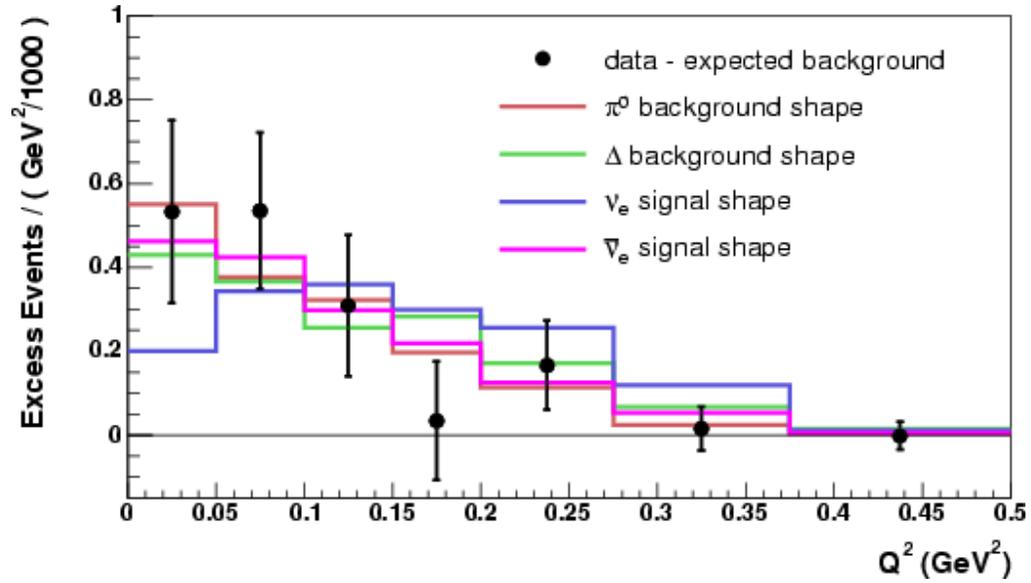}}
\caption{\em The neutrino event excess as a
function of $Q^2$ for $300 < E_\nu^{QE} < 475$ MeV.}
\label{data_mc4}
\end{figure}

\begin{figure}
\centerline{\includegraphics[height=3.in]{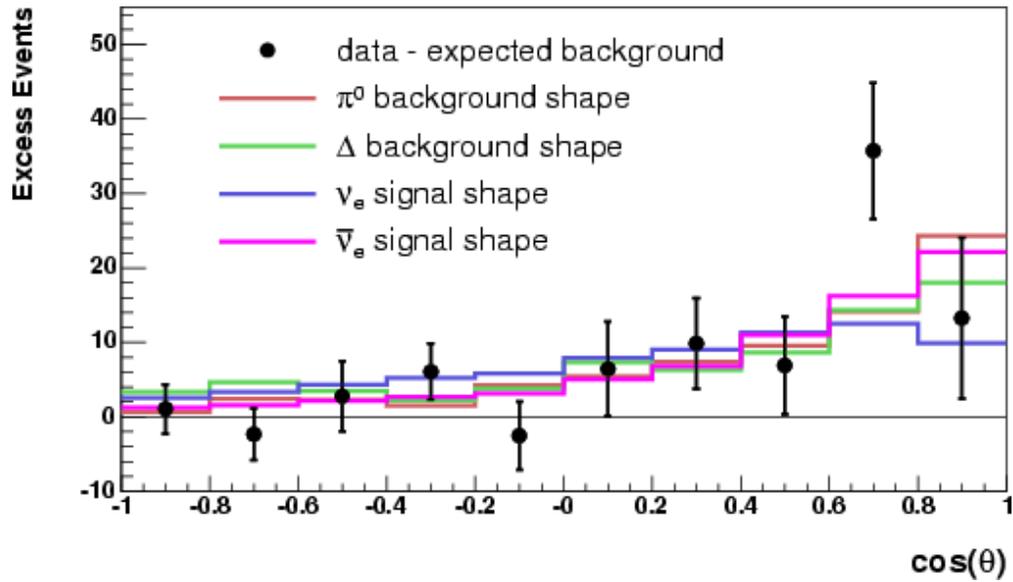}}
\caption{\em The neutrino event excess as a
function of $\cos (\theta)$ for $300 < E_\nu^{QE} < 475$ MeV.}
\label{data_mc5}
\end{figure}

Figs. \ref{data_mc4} and \ref{data_mc5} show the event excess as
functions of $Q^2$ and $\cos (\theta)$ for events in the
$300 < E_\nu^{QE} < 475$ MeV range, where
$Q^2$ is determined from the energy and angle of the outgoing lepton and
$\theta$ is the angle between the beam direction and the
reconstructed event direction.
Also shown in the figures are the
expected shapes from $\nu_e C \rightarrow e^- X$ and
$\bar \nu_e C \rightarrow e^+ X$ charged-current (CC) scattering and from
the NC $\pi^0$ and $\Delta \rightarrow N \gamma$
reactions, which are representative of photon events
produced by NC scattering.
The NC scattering assumes the $\nu_\mu$ energy spectrum, while the CC
scattering assumes the transmutation of $\nu_\mu$ into $\nu_e$ and
$\bar \nu_e$, respectively.
As shown in Table \ref{chisquare}, the $\chi^2$ values from
comparisons of the event excess to the expected shapes are acceptable for all
of the processes. However, any of the backgrounds in Table \ref{chisquare} 
would have to be increased by $>5 \sigma$ to explain the low-energy excess.

\begin{table}
\begin{center}
\caption{\label{chisquare} \em The
$\chi^2$ values
from comparisons of the neutrino
event excess $Q^2$ and $\cos (\theta)$ distributions
for $300 < E_\nu^{QE} < 475$ MeV
to the expected shapes from
various NC and CC reactions. Also shown is the factor increase necessary
for the estimated background for each process
to explain the low-energy excess and the corresponding number of sigma.}
\vspace{0.1in}
\begin{tabular}{|c|c|c|c|}
\hline
Process&$\chi^2(cos \theta)/9$ DF&$\chi^2(Q^2)/6$ DF&Factor Increase \\
\hline
NC $\pi^0$&13.46&2.18&2.0 $(6.8 \sigma)$ \\
$\Delta \rightarrow N \gamma$&16.85&4.46&2.7 $(18.4 \sigma)$ \\
$\nu_e C \rightarrow e^- X$&14.58&8.72&2.4 $(15.3 \sigma)$ \\
$\bar \nu_e C \rightarrow e^+ X$&10.11&2.44&65.4 $(41.0 \sigma)$ \\
\hline
\end{tabular}
\end{center}
\end{table}

\subsection{Initial Antineutrino Oscillation Results}

\begin{figure}
\centering
\includegraphics[width=10.5cm,
scale=1.0]{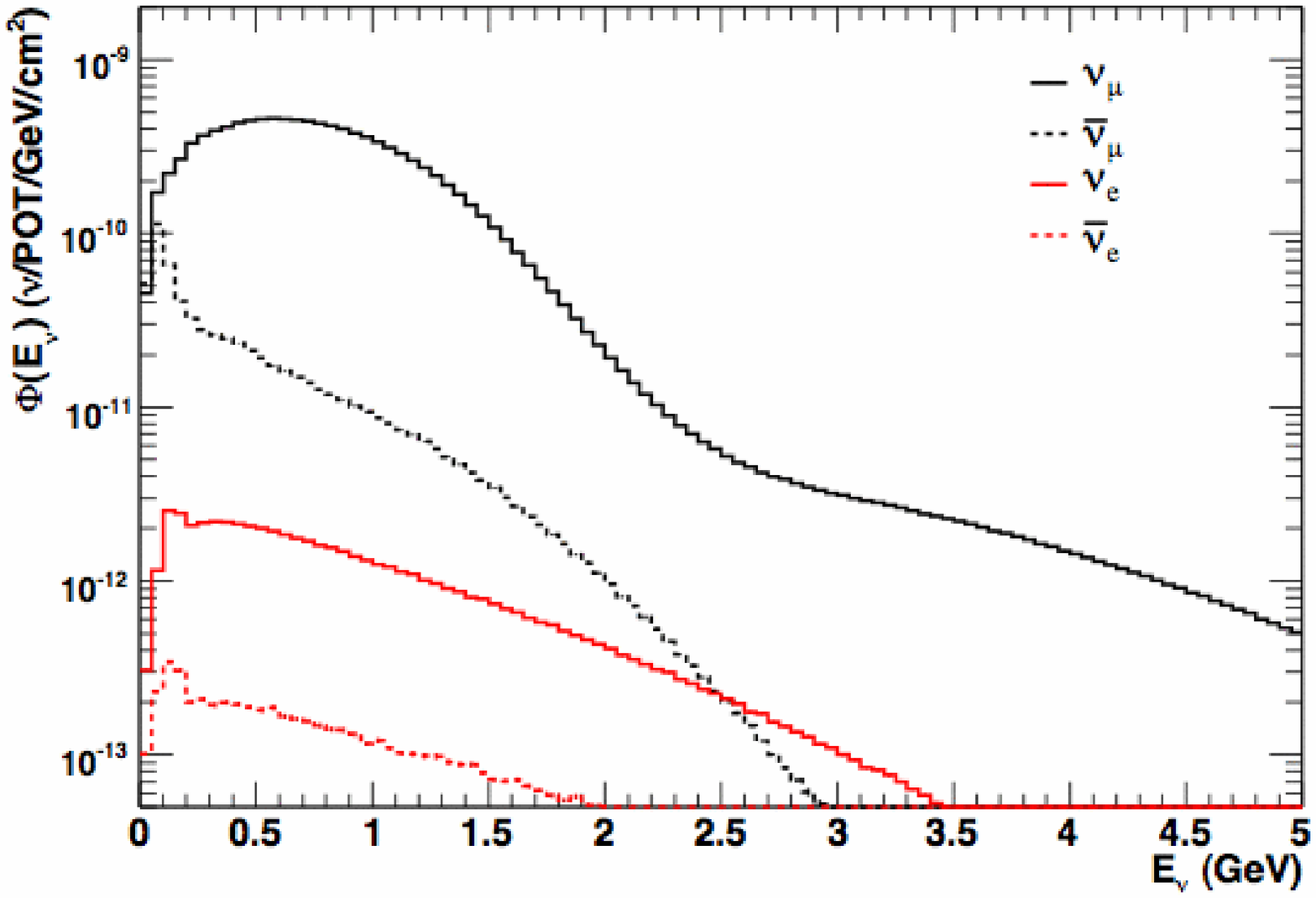}
\hspace{1cm}
\includegraphics[width=10.5cm,
scale=1.0]{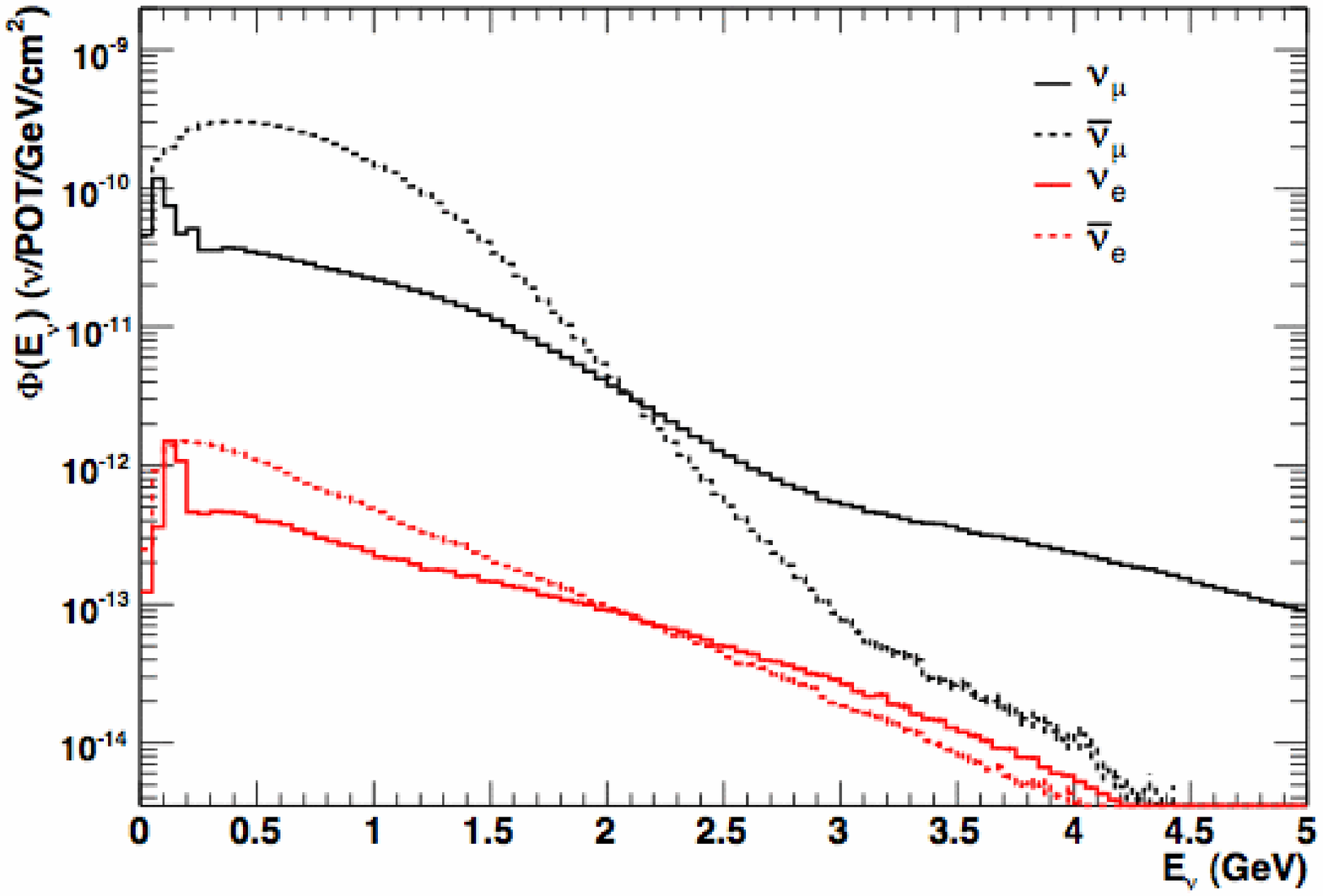}
\parbox{7in}{
\caption{\em The estimated neutrino fluxes for neutrino mode (top plot) and
antineutrino mode (bottom plot).}
\label{nu_flux}}
\end{figure}

\begin{figure}
\centering
\includegraphics[width=7.5cm,
scale=1.0]{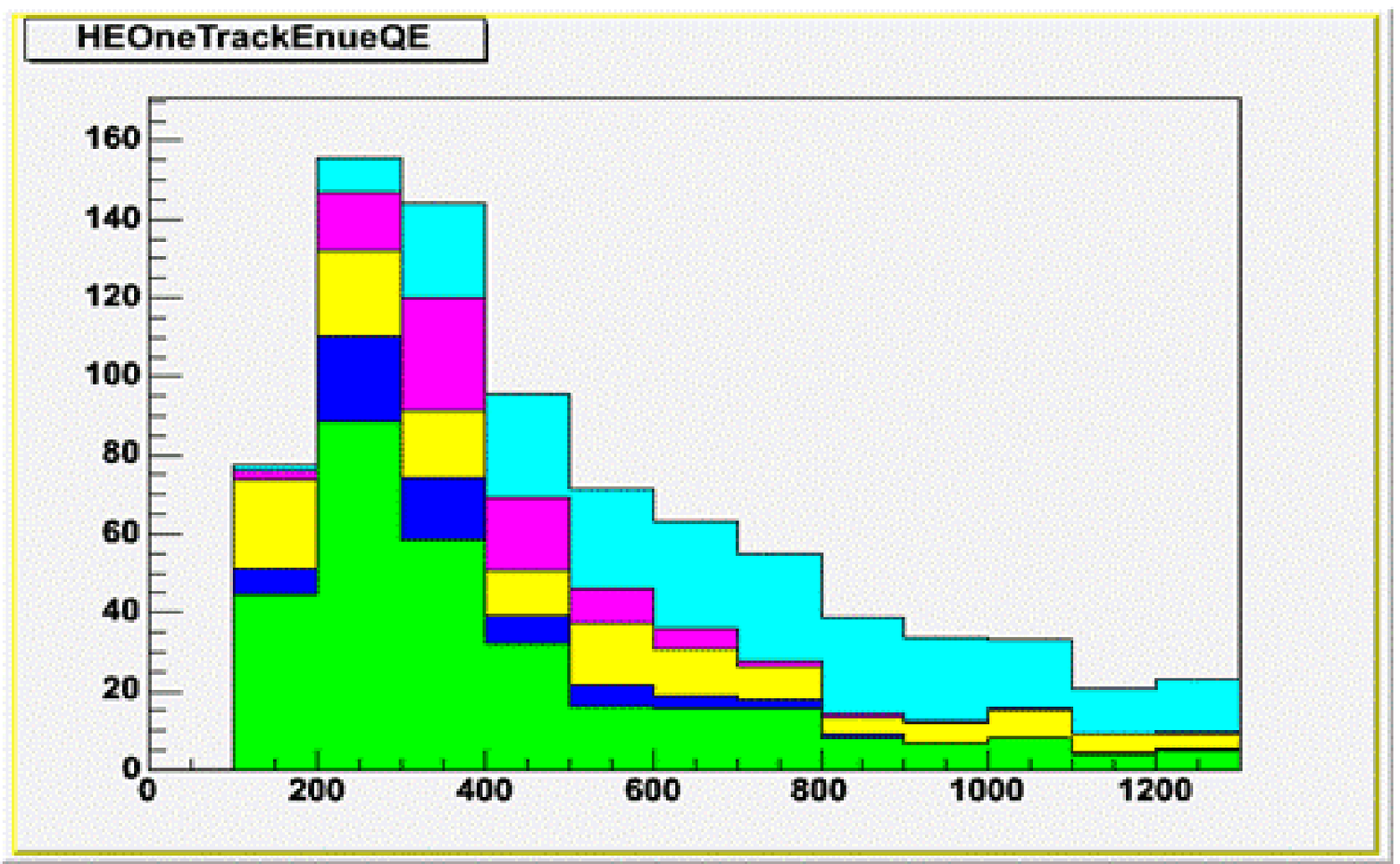}
\hspace{1cm}
\includegraphics[width=7.5cm,
scale=1.0]{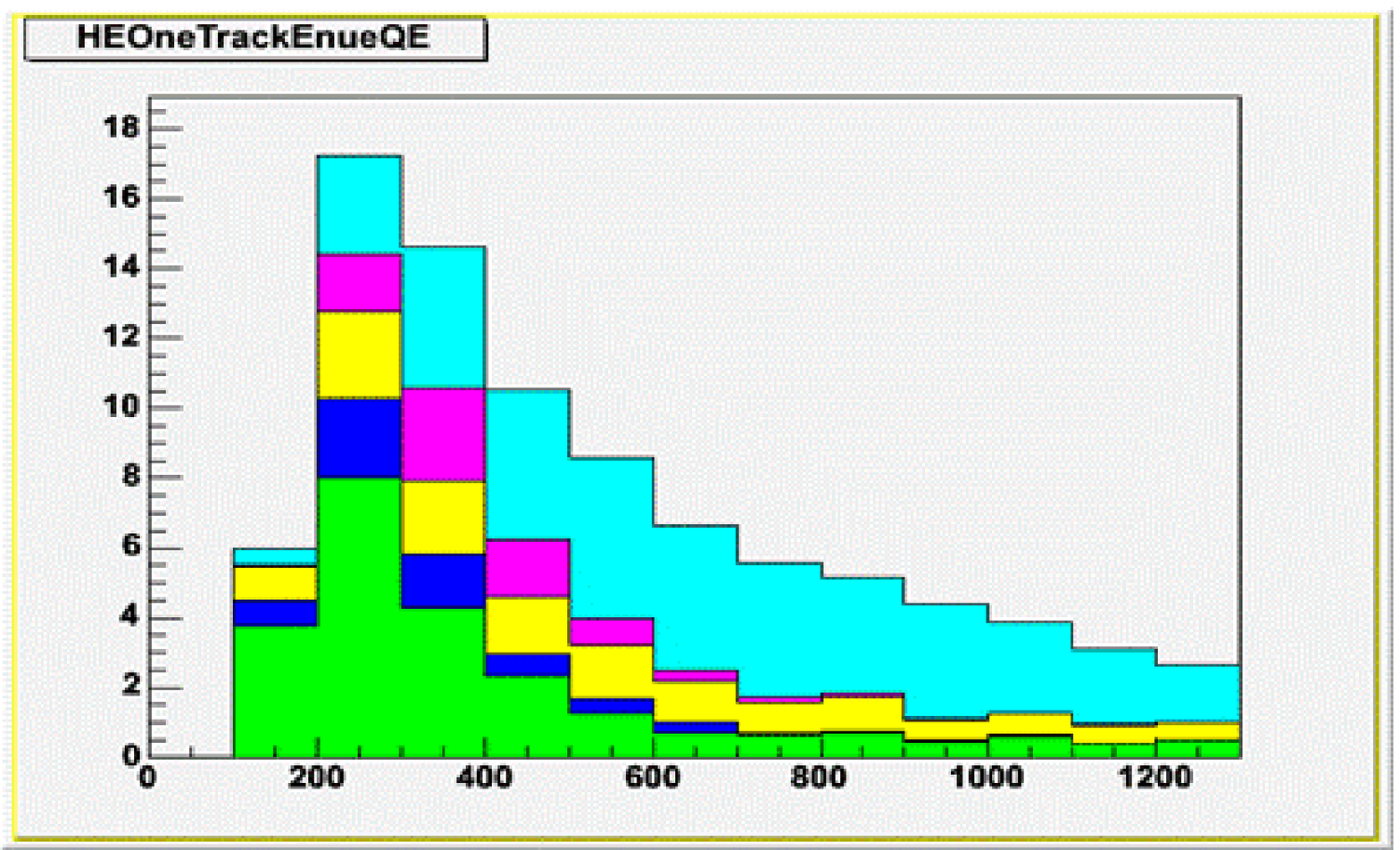}
\parbox{7in}{
\caption{\em The estimated backgrounds for the neutrino oscillation
search in neutrino mode (top plot) and antineutrino mode (bottom plot).
The $\pi^0$, $\Delta \rightarrow N \gamma$,
intrinsic $\nu_e$/$\bar \nu_e$, external event, and other backgrounds
correspond to the green, pink, light blue, blue, and yellow colors,
respectively.}
\label{nu_bkgd}}
\end{figure}

\begin{figure}
\centerline{\includegraphics[height=3.5in]{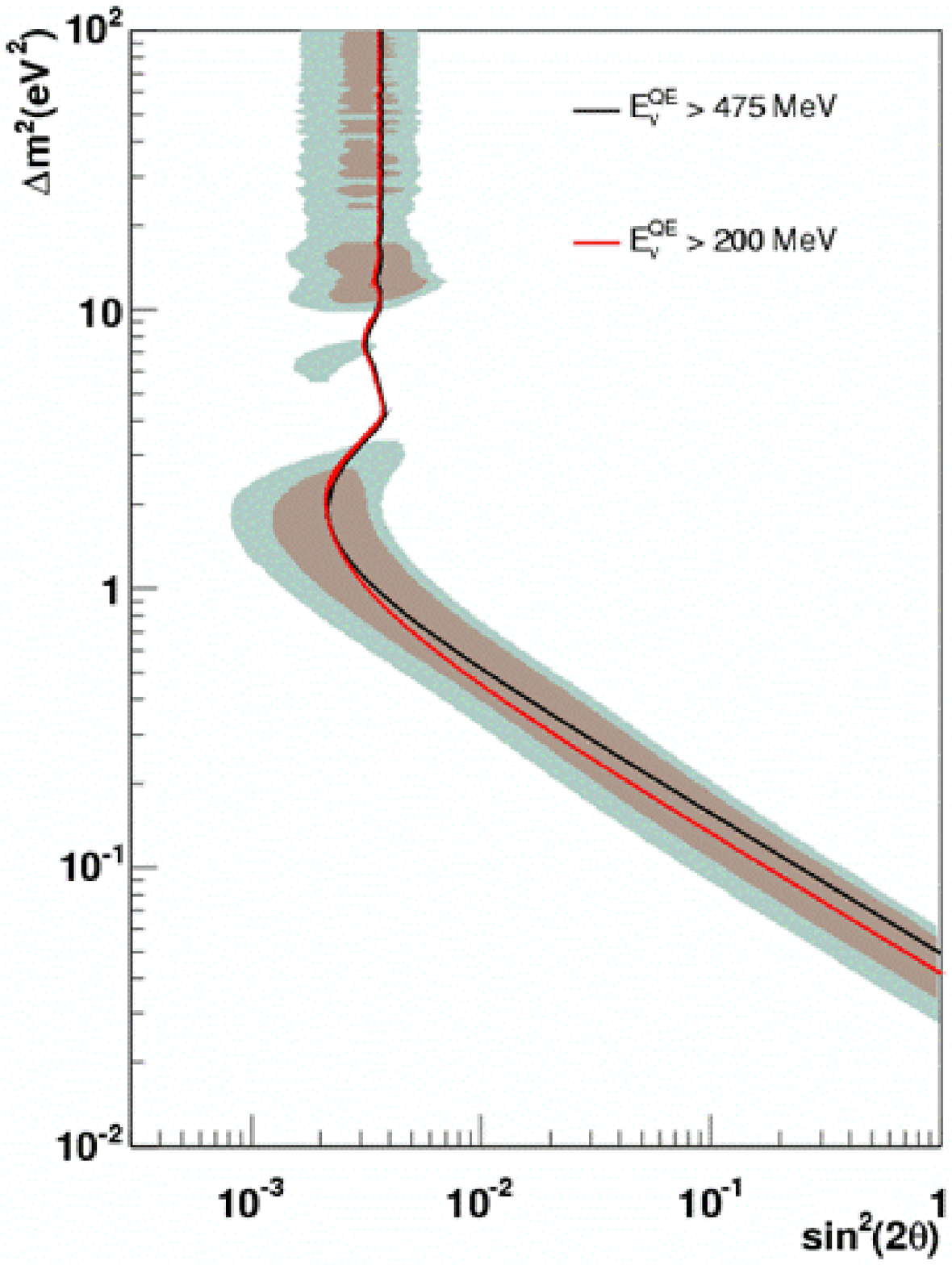}}
\caption{\em The expected antineutrino oscillation
sensitivity at 90\% CL for the present data sample corresponding to 3.4E20 POT.
The
two sensitivity
curves correspond to threshold energies of 200 MeV (red curve) and 475 MeV
(black curve).}
\label{antinu_sens}
\end{figure}

The same analysis that was used for the neutrino oscillation results
is employed for the initial antineutrino oscillation results \cite{mb_anti}.
Fig. \ref{nu_flux} shows the estimated neutrino fluxes for neutrino
mode and antineutrino mode, respectively. The fluxes are fairly similar
(the intrinsic electron-neutrino background is approximately 0.5\% for
both modes of running), although the wrong-sign contribution to the
flux in antineutrino mode ($\sim 18\%$) is much larger than in neutrino mode
($\sim 6\%$). The average $\nu_e$ plus $\bar \nu_e$
energies are 0.96 GeV in neutrino mode and 0.77 GeV in antineutrino mode,
while the average $\nu_\mu$ plus $\bar \nu_\mu$ energies are
0.79 GeV in neutrino mode and 0.66 GeV in antineutrino mode.
Also, as shown in Fig. \ref{nu_bkgd}, the
estimated backgrounds in the two modes are very similar, especially at low energy.
Fig. \ref{antinu_sens} shows the expected antineutrino oscillation
sensitivity for the present data sample corresponding to 3.4E20 POT. The
two sensitivity
curves correspond to threshold neutrino energies of 200 MeV and 475 MeV.

\begin{figure}
\centerline{\includegraphics[height=3.5in]{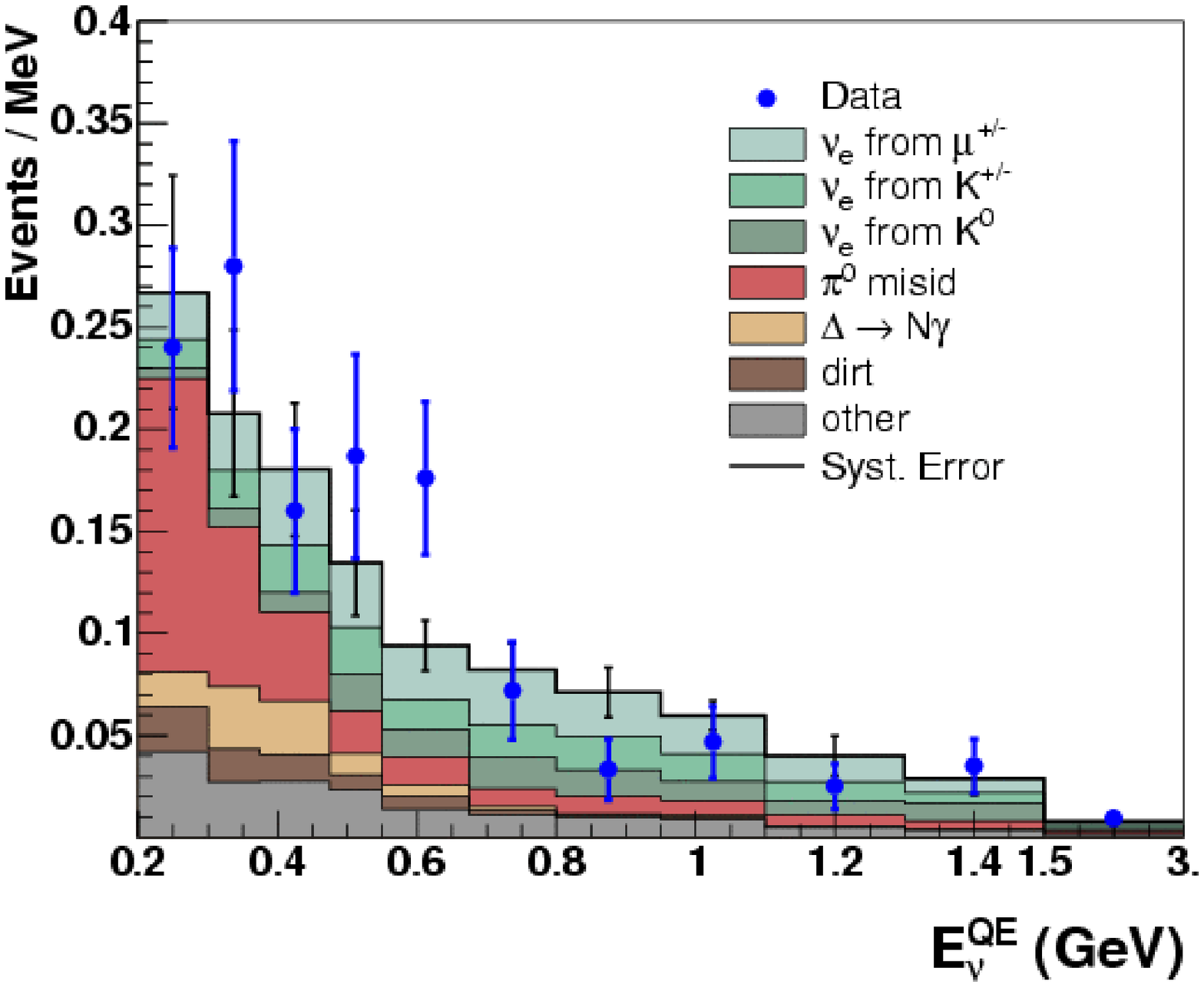}}
\caption{\em The MiniBooNE reconstructed antineutrino energy distribution
for candidate $\bar \nu_e$ data events (points with error bars) compared to
the Monte Carlo simulation (histogram).}
\label{antinu_data_bkgd}
\end{figure}

\begin{figure}
\centerline{\includegraphics[height=5.5in]{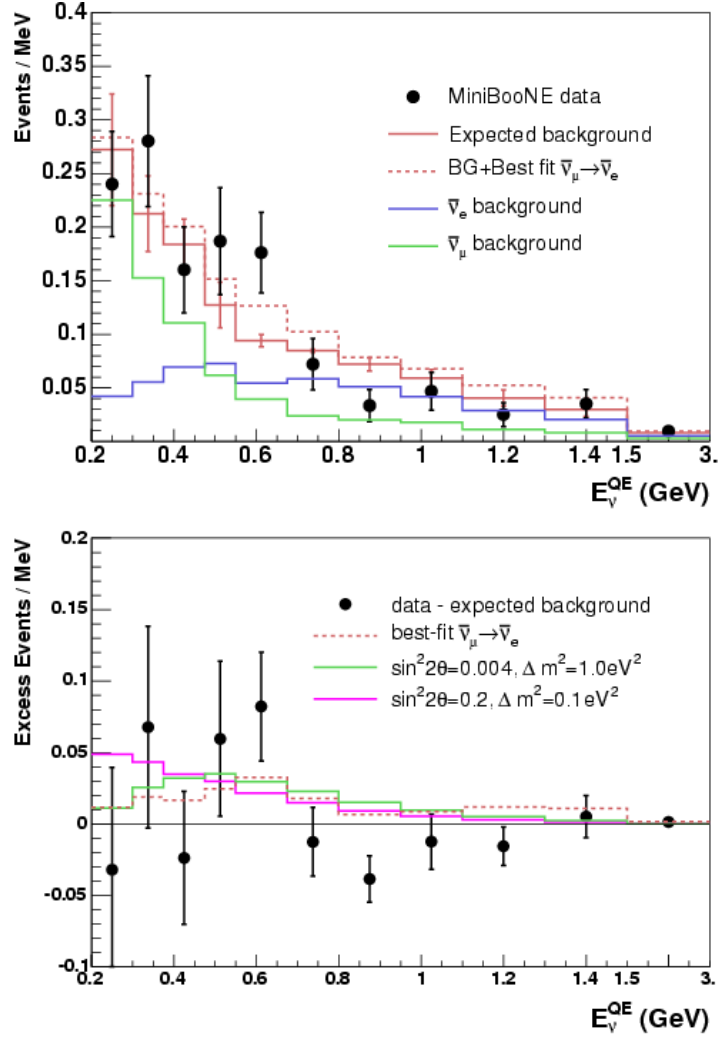}}
\caption{\em The MiniBooNE reconstructed antineutrino energy distribution
for candidate $\bar \nu_e$ data events (top) and the
excess number of events (bottom)
as a function of reconstructed neutrino energy for the present antineutrino
data sample corresponding to 3.4E20 POT. Also shown are the expectations from
the best oscillation fit and from oscillation parameters in the LSND allowed region.}
\label{antinu_excess1}
\end{figure}

\begin{figure}
%\centerline{\includegraphics[height=5.5in,angle=-90]{evis_constr_uncertainties.eps}}
\centerline{\includegraphics[height=5.5in,angle=-90]{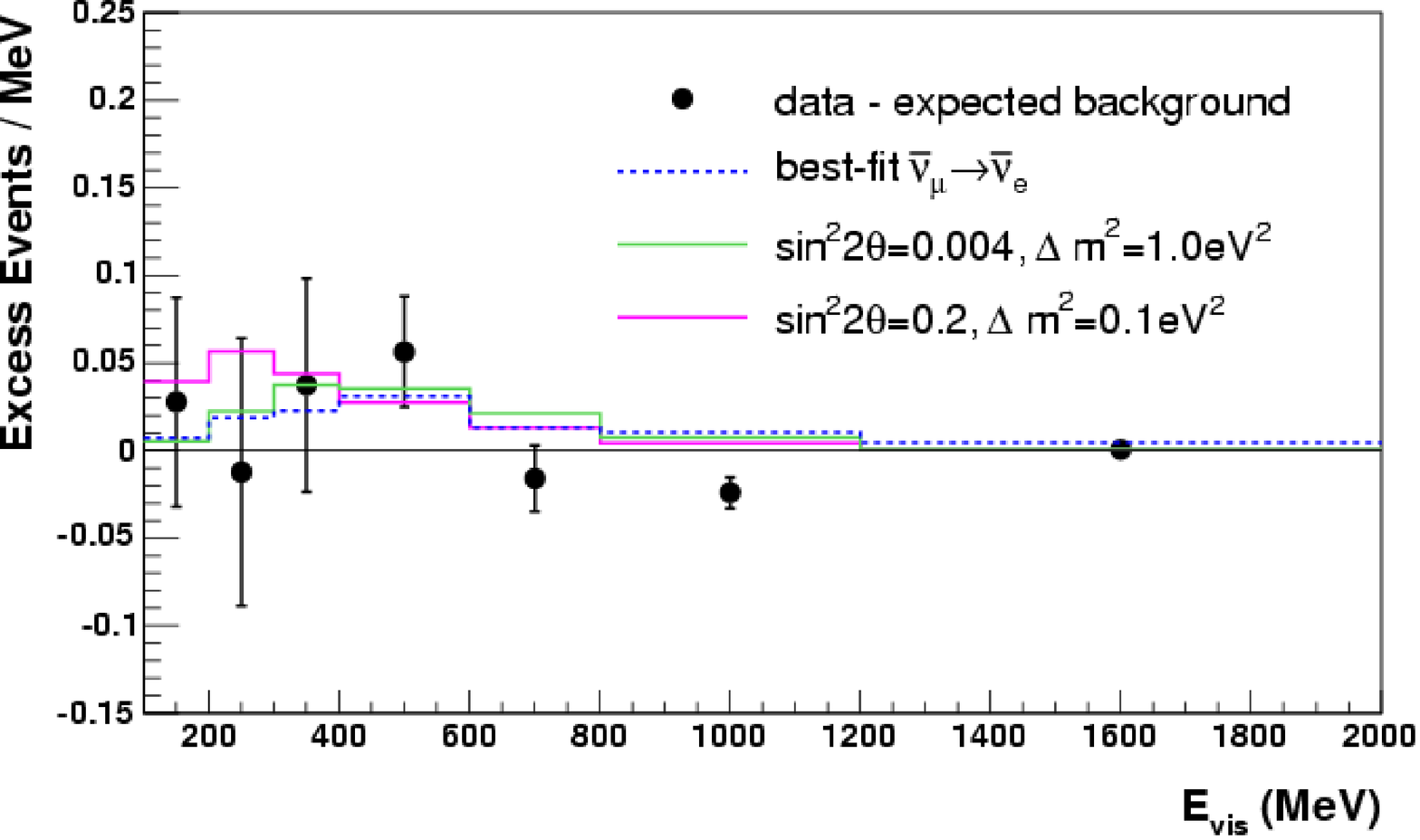}}
\caption{\em The excess number of candidate $\bar \nu_e$
events (data minus Monte Carlo expectation)
as a function of visible energy for the present antineutrino data sample corresponding to 3.4E20 POT.
Also shown are the expectations from
the best oscillation fit and from oscillation parameters in the LSND allowed region.}
\label{antinu_excess2}
\end{figure}

The initial oscillation results for antineutrino mode are shown
in Table \ref{antinu_stat} and Figs. \ref{antinu_data_bkgd} through
\ref{antinu_excess2}. It is remarkable that no excess ($-0.5 \pm 7.8 \pm 8.7$
events) is observed in the low-energy range $200<E^{QE}_{\nu}<475$ MeV. In 
order to understand the implications that the antineutrino data have on 
the neutrino
low-energy excess, Table \ref{implications} shows the expected excess
of low-energy events in antineutrino mode under various hypotheses.
These hypotheses include the following:

\begin{itemize}
\item Same $\sigma$: Same cross section for neutrinos and antineutrinos.
\item $\pi^0$ Scaled: Scaled to number of neutral-current $\pi^0$ events.
\item POT Scaled: Scaled to number of POT.
\item BKGD Scaled: Scaled to total background events.
\item CC Scaled: Scaled to number of charged-current events.
\item Kaon Scaled: Scaled to number of low-energy kaon events.
\item Neutrino Scaled: Scaled to number of neutrino events.
\end{itemize}

Also shown in the Table is the probability (from a two-parameter fit
to the data) that each hypothesis explains the observed number of low-energy
neutrino and antineutrino events, assuming only statistical errors, correlated
systematic errors, and uncorrelated systematic errors. A proper treatment of
the systematic errors is in progress; however, it is clear from the Table that
the ``Neutrino Scaled'' hypothesis fits best and that
the ``Same $\sigma$'', ``POT Scaled'', and ``Kaon Scaled''
hypotheses are strongly disfavored. It will be very
important to understand this unexpected difference between neutrino and
antineutrino properties.

\begin{table}
%\begin{table}[h]
\begin{center}
\caption{\label{antinu_stat} \em The number of antineutrino data,
background, and excess events for different
$E_{\bar \nu}^{QE}$ ranges, together with the
significance of the excesses in antineutrino mode.}
%%\begin{ruledtabular}
\vspace{0.1in}
\begin{tabular}{|c|c|}
\hline
Event Sample&Final Analysis \\
\hline
$200-475$ MeV& \\
Data&61 \\
Background&$61.5 \pm 7.8 \pm 8.7$ \\
Excess&$-0.5 \pm 7.8 \pm 8.7$ \\
Significance&$-0.04 \sigma$ \\
\hline
$475-1250$ MeV& \\
Data&61 \\
Background&$57.8 \pm 7.6 \pm 6.5$ \\
Excess&$3.2 \pm 7.6 \pm 6.5$ \\
Significance&$0.3 \sigma$ \\
\hline
$475-3000$ MeV& \\
Data&83 \\
Background&$77.4 \pm 8.8 \pm 9.6$ \\
Excess&$5.6 \pm 8.8 \pm 9.6$ \\
Significance&$0.4 \sigma$ \\
\hline
\end{tabular}
%\end{center}
%%\end{ruledtabular}
\end{center}
\end{table}

\begin{table}
%\begin{table}[h]
\begin{center}
\caption{\label{implications} \em The expected excess
of low-energy events in antineutrino mode under various hypotheses for 3.4E20 POT.
Also shown in the Table is the probability (from a two-parameter fit
to the data) that each hypothesis explains the observed number of low-energy
neutrino and antineutrino events, assuming only statistical errors, correlated
systematic errors, and uncorrelated systematic errors.}
%%\begin{ruledtabular}
\vspace{0.1in}
\begin{tabular}{|c|c|c|c|c|}
\hline
Hypothesis&\# of $\bar \nu$ Events&Stat. Err.&Cor. Syst. Err.&Uncor. Syst. Err.
\\
\hline
Same $\sigma$&37.2&0.1\%&0.1\%&6.7\% \\
$\pi^0$ Scaled&19.4&3.6\%&6.4\%&21.5\% \\
POT Scaled&67.5&0.0\%&0.0\%&1.8\% \\
BKGD Scaled&20.9&2.7\%&4.7\%&19.2\% \\
CC Scaled&20.4&2.9\%&5.2\%&19.9\% \\
Kaon Scaled&39.7&0.1\%&0.1\%&5.9\% \\
Neutrino Scaled&6.7&38.4\%&51.4\%&58.0\% \\
\hline
\end{tabular}
%\end{center}
%%\end{ruledtabular}
\end{center}
\end{table}

\begin{figure}
\centerline{\includegraphics[height=4.5in]{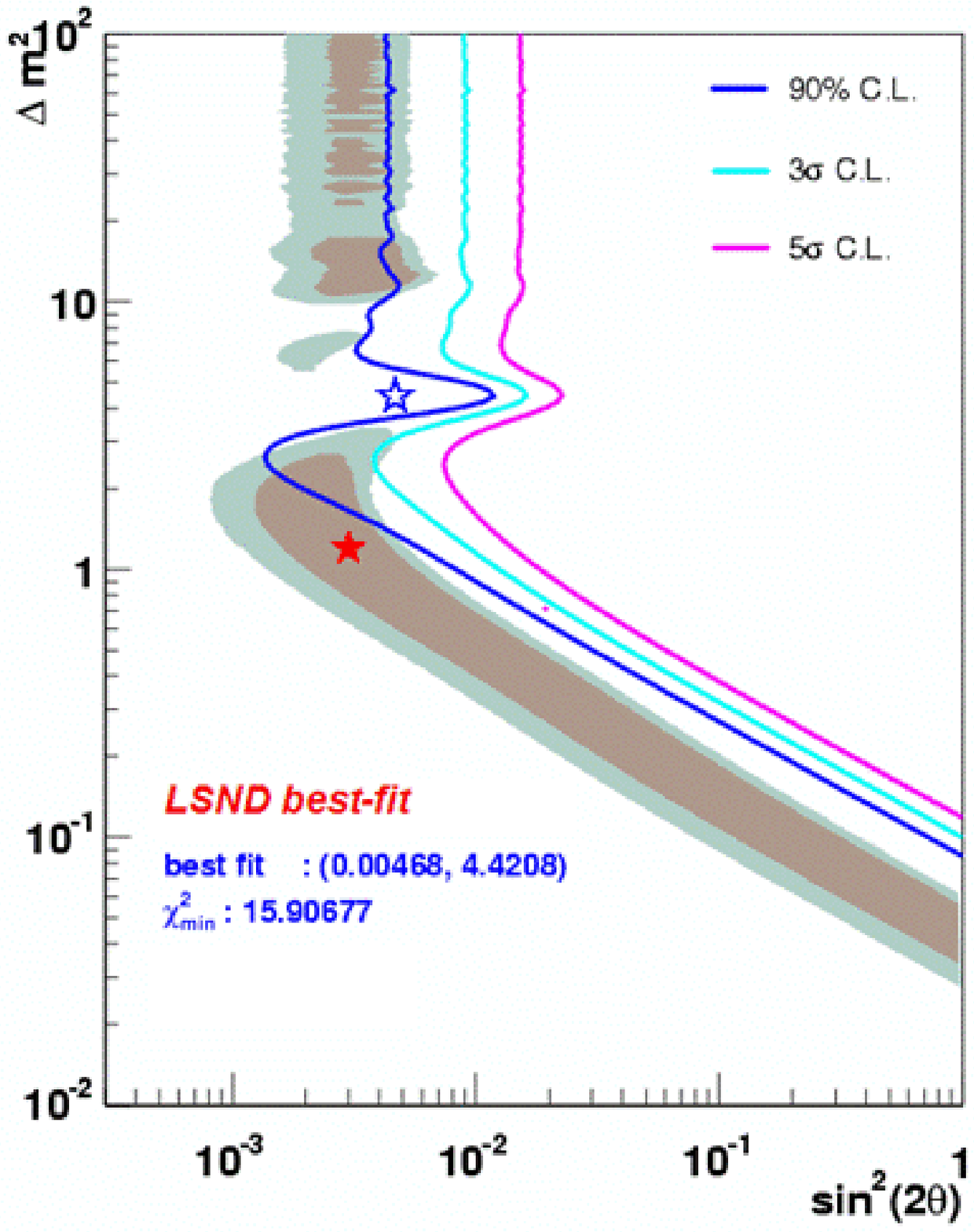}}
\caption{\em The antineutrino oscillation allowed region in the energy range
$475<E_{\bar \nu}^{QE}<3000$ MeV for the present antineutrino
data sample corresponding to 3.4E20 POT. Also shown are the best oscillation fit
($\Delta m^2 = 4.4$ eV$^2$, $\sin^22\theta = 0.0047$, corresponding
to an excess of $18.6 \pm 13.2$
events) and the LSND best-fit point ($\Delta m^2 = 1.2$ eV$^2$,
$\sin^22\theta = 0.003$, corresponding to an excess of $14.7$ events).}
\label{antinu_allowed2}
\end{figure}

The antineutrino data were also fit for oscillations in the energy range
$475<E_{\bar \nu}^{QE}<3000$ MeV, assuming antineutrino oscillations but no
neutrino oscillations. The antineutrino oscillation allowed region is
shown in Fig. \ref{antinu_allowed2}. At present, the oscillation limit is worse
than
the sensitivity. The best oscillation fit corresponds to
$\Delta m^2 = 4.4$ eV$^2$, $\sin^22\theta = 0.0047$, and a fitted excess of $18.6 \pm 13.2$
events, which is consistent with the LSND best-fit point of $\Delta m^2 = 1.2$ eV$^2$,
$\sin^22\theta = 0.003$, and an expected excess of $14.7$ events. With the present
antineutrino statistics, the data are consistent with both the LSND best-fit point
and the null point, although the LSND best-fit point has a better $\chi^2$
($\chi^2 = 17.63/15$ DF, probability = 30\%) than the null point
($\chi^2 = 22.19/15$ DF, probability = 10\%).

\subsection{MiniBooNE NuMI Results}

\begin{figure}
\centerline{\includegraphics[height=4.0in]{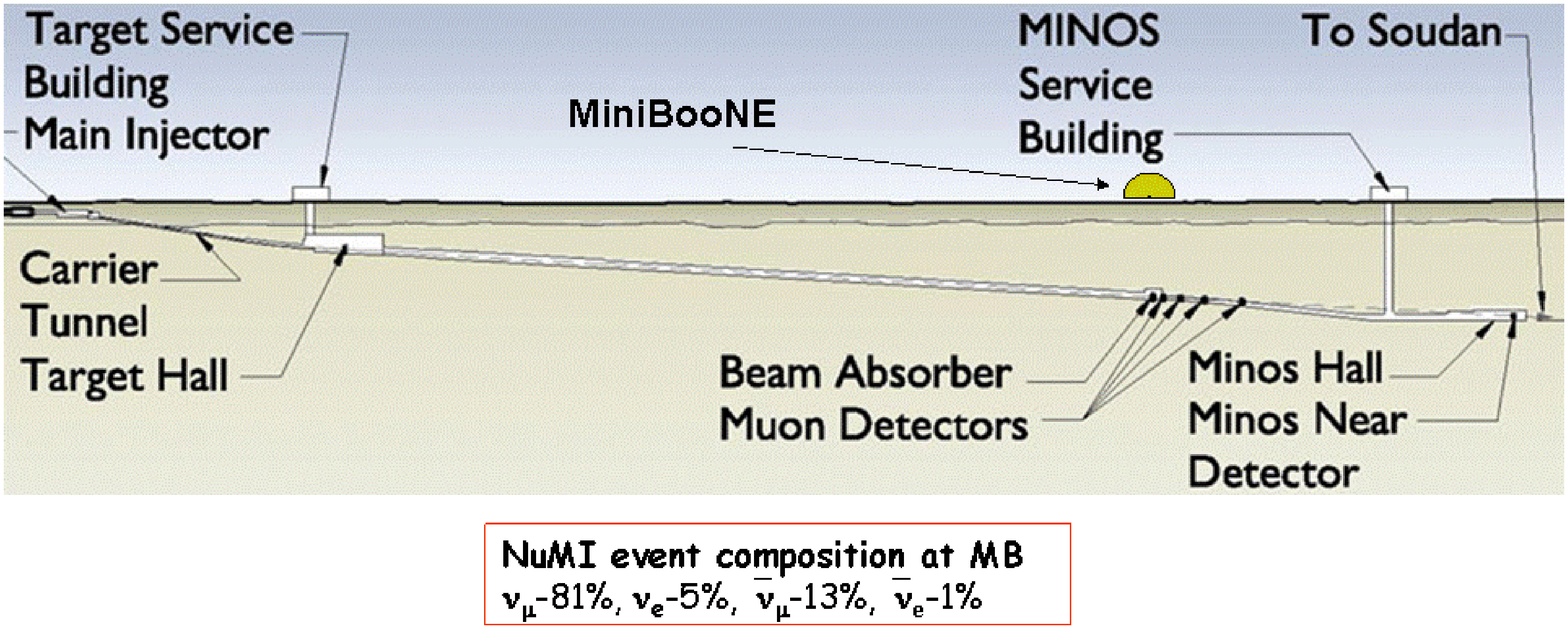}}
%\centerline{\psfig{file=numi.eps,width=16.cm}}
%\includegraphics[width=40pc]{numi.eps}
\caption{\label{numi} \em The NuMI beam.}
\end{figure}

\begin{figure}
\centerline{\includegraphics[height=3.0in]{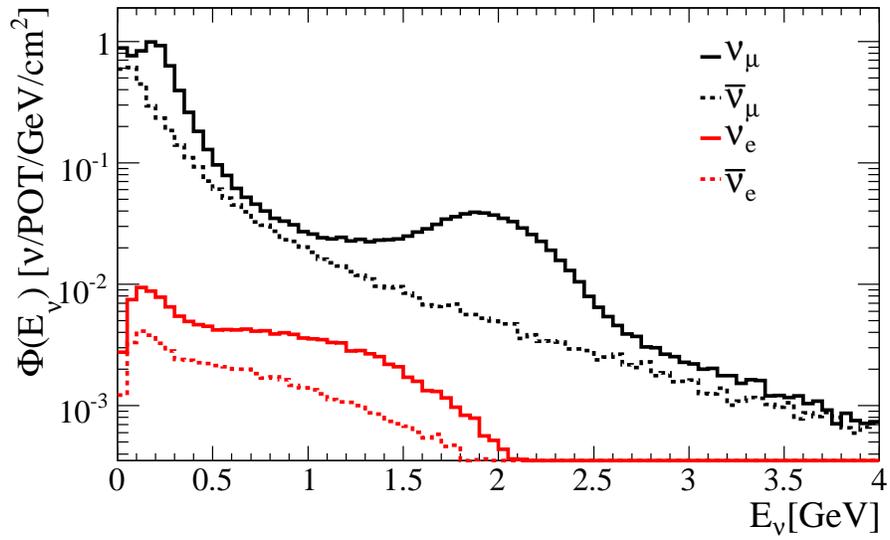}}
%\centerline{\psfig{file=numi.eps,width=16.cm}}
%\includegraphics[width=40pc]{numi.eps}
\caption{\label{numi_flux} \em The estimated neutrino flux at the MiniBooNE
detector from the NuMI beam.}
\end{figure}

\begin{figure}
\centerline{\includegraphics[height=3.0in]{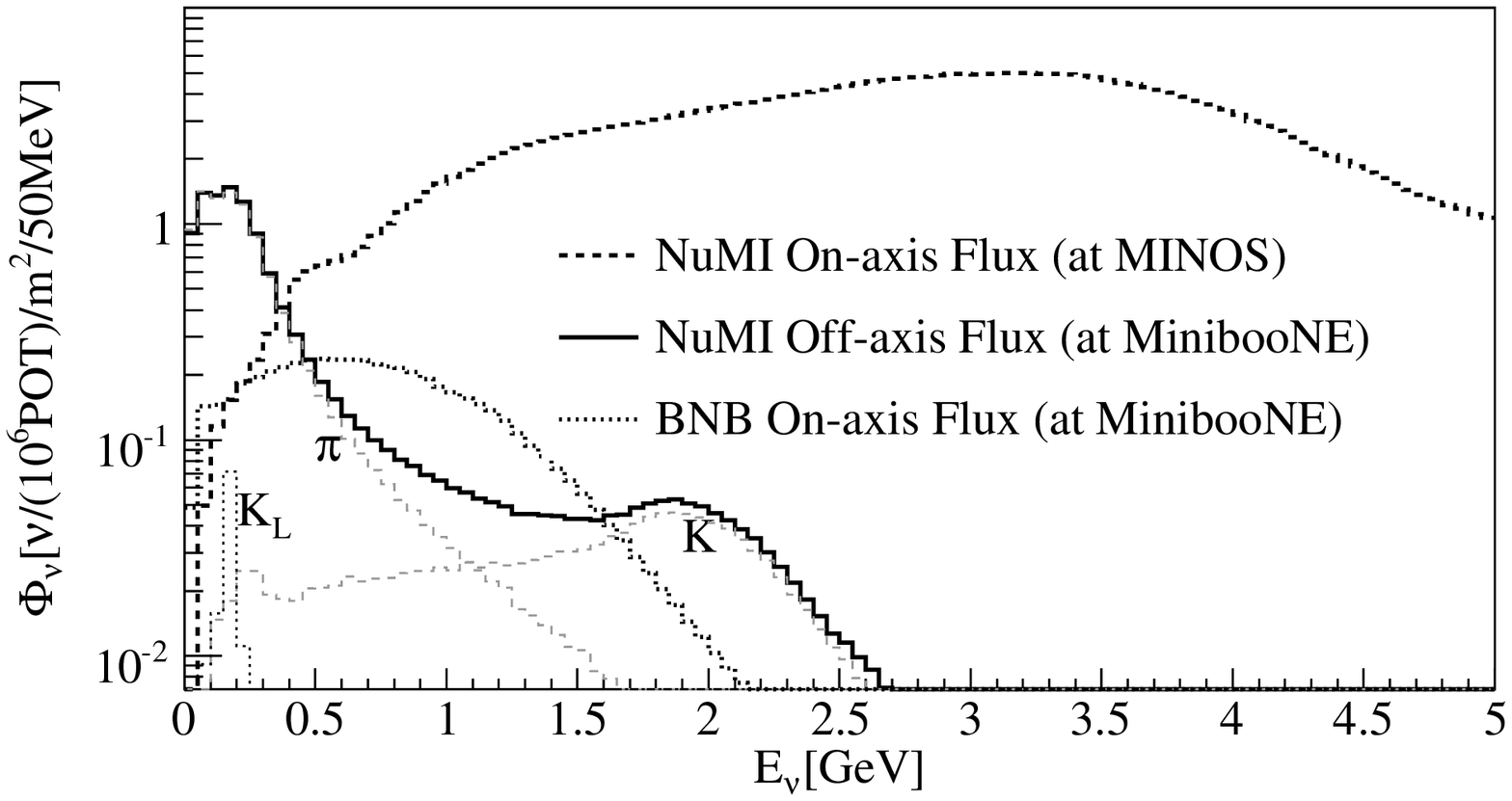}}
%\centerline{\psfig{file=numi.eps,width=16.cm}}
%\includegraphics[width=40pc]{numi.eps}
\caption{\label{numi_bnb} \em A comparison between the BNB and NuMI neutrino fluxes.}
\end{figure}

Neutrino events are also observed in MiniBooNE from
the NuMI beam \cite{mb_numi}. The NuMI beam,
as shown in Fig. \ref{numi}, differs
from the Booster neutrino beam (BNB) in several respects.
First, the NuMI beam is off axis by 110 mrad, whereas the BNB is on axis.
Second, neutrinos from NuMI travel $\sim 700$ m, compared
to $\sim 500$ m for neutrinos from the BNB. Also, the NuMI beam has a
6\% contribution from electron-neutrinos and a 14\% contribution from
antineutrinos, while the BNB percentages are $0.5\%$ and $2\%$, respectively.
Fig. \ref{numi_flux} shows the estimated neutrino flux at the MiniBooNE
detector from the NuMI beam, while Fig. \ref{numi_bnb} compares the neutrino
fluxes from the BNB and NuMI beams.

\begin{figure}
\centerline{\includegraphics[height=3.0in]{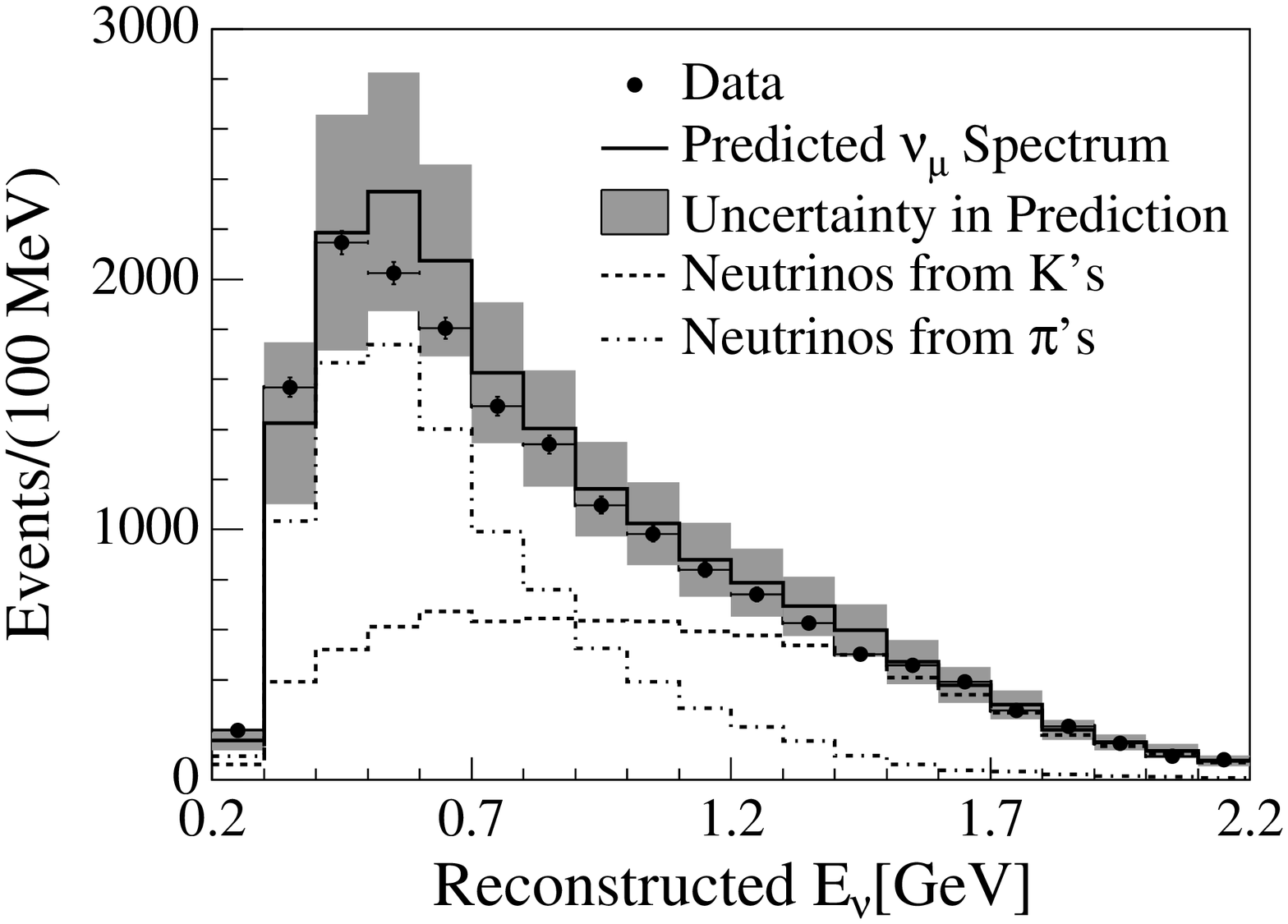}}
%\centerline{\psfig{file=figure_3_numi.eps,width=14.cm}}
\caption{\label{numi_excess_mu} \em The comparison between data events (points
with error bars) and the MC simulation (histogram) for
NuMI-induced $\nu_\mu$
CCQE candidate events.}
\end{figure}

\begin{figure}
\centerline{\includegraphics[height=3.0in]{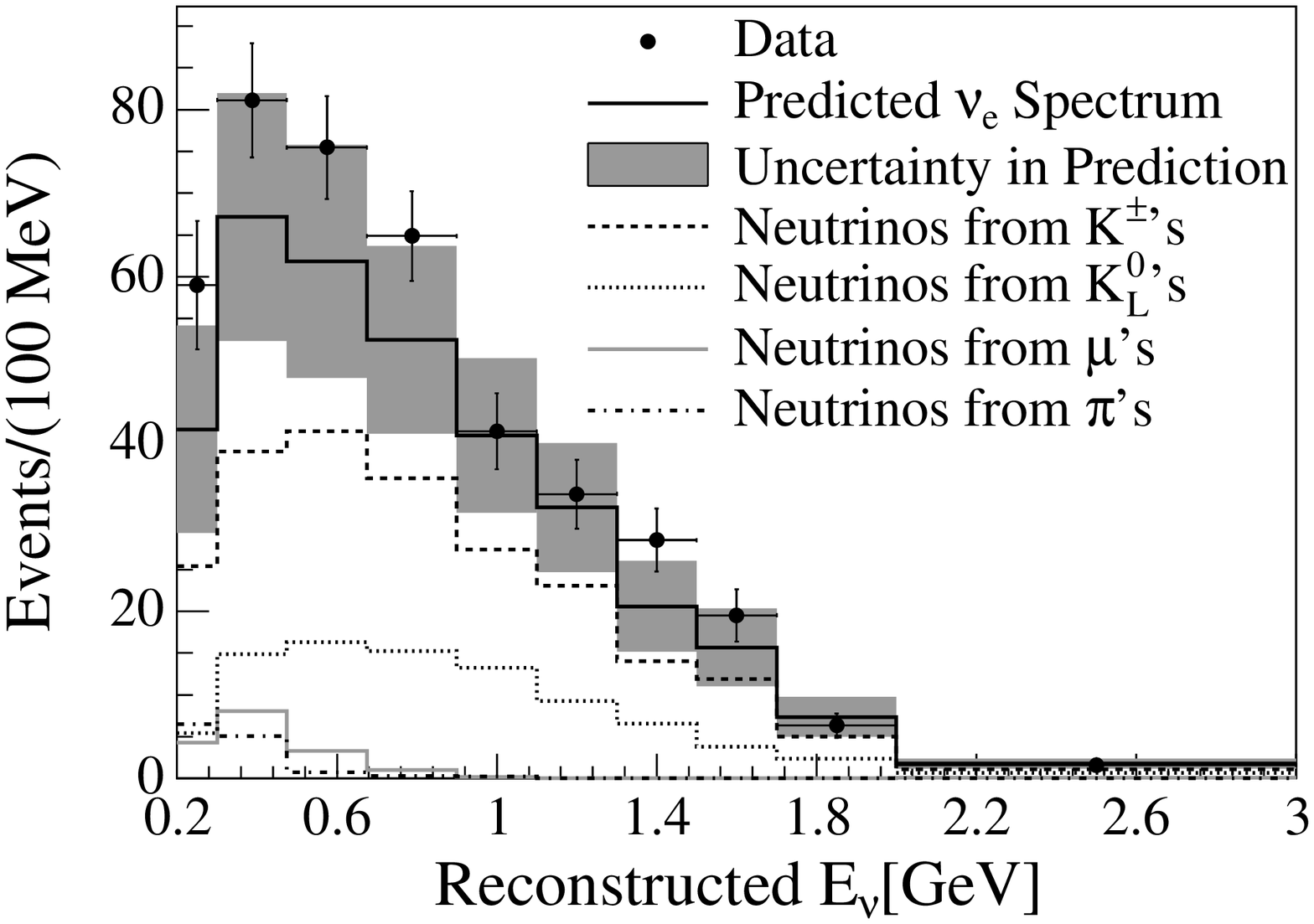}}
%\centerline{\psfig{file=figure_5_numi.eps,width=14.cm}}
\caption{\label{numi_excess_e} \em The comparison between data events (points
with error bars) and the MC simulation (histogram) for
NuMI-induced $\nu_e$
CCQE candidate events.}
\end{figure}

Figs. \ref{numi_excess_mu} and \ref{numi_excess_e}
show the comparison between data events (points
with error bars) and the MC simulation (histogram) for $\nu_\mu$
CCQE candidate events and $\nu_e$ CCQE candidate events, respectively.
Although the systematic errors are presently large, the data are
observed to be systematically low for $\nu_\mu$
CCQE candidate events and systematically high for $\nu_e$
CCQE candidate events. Updated results should be available soon
with three times the data sample and with reduced systematic errors
by constraining the normalization to the $\nu_\mu$ sample.

The NuMI data analysis
is currently directed toward examining the low-energy region and searching
for neutrino oscillations.
This will complement the analysis done with MiniBooNE using
neutrino and anti-neutrino BNB data, but with different systematic errors.
It is worth noting that the NuMI $\nu_{e}$ CCQE sample has a very
different composition
when compared to the BNB neutrino $\nu_{e}$ CCQE sample. The BNB $\nu_{e}$
CCQE sample
originates mostly from decays of pions and muons and
contains a large fraction
of $\nu_{\mu}$-induced mis-identified events. On other hand, the NuMI $\nu_{e}$
CCQE sample is produced
mostly from the decay of kaons and contains a dominant fraction of intrinsic
$\nu_{e}$ events.
The analysis will be done by forming a correlation between the
$\nu_{\mu}$ CCQE
and $\nu_{e}$ CCQE samples and by tuning the prediction to the data
simultaneously. The result is that
common systematics cancel, and this might reveal something
important about the nature
of the $\nu_e$ sample.

\subsection{MiniBooNE Disappearance Results}

\begin{figure}
%\centerline{\includegraphics[height=6in]{3+1.eps}}
\centerline{\includegraphics[height=6in]{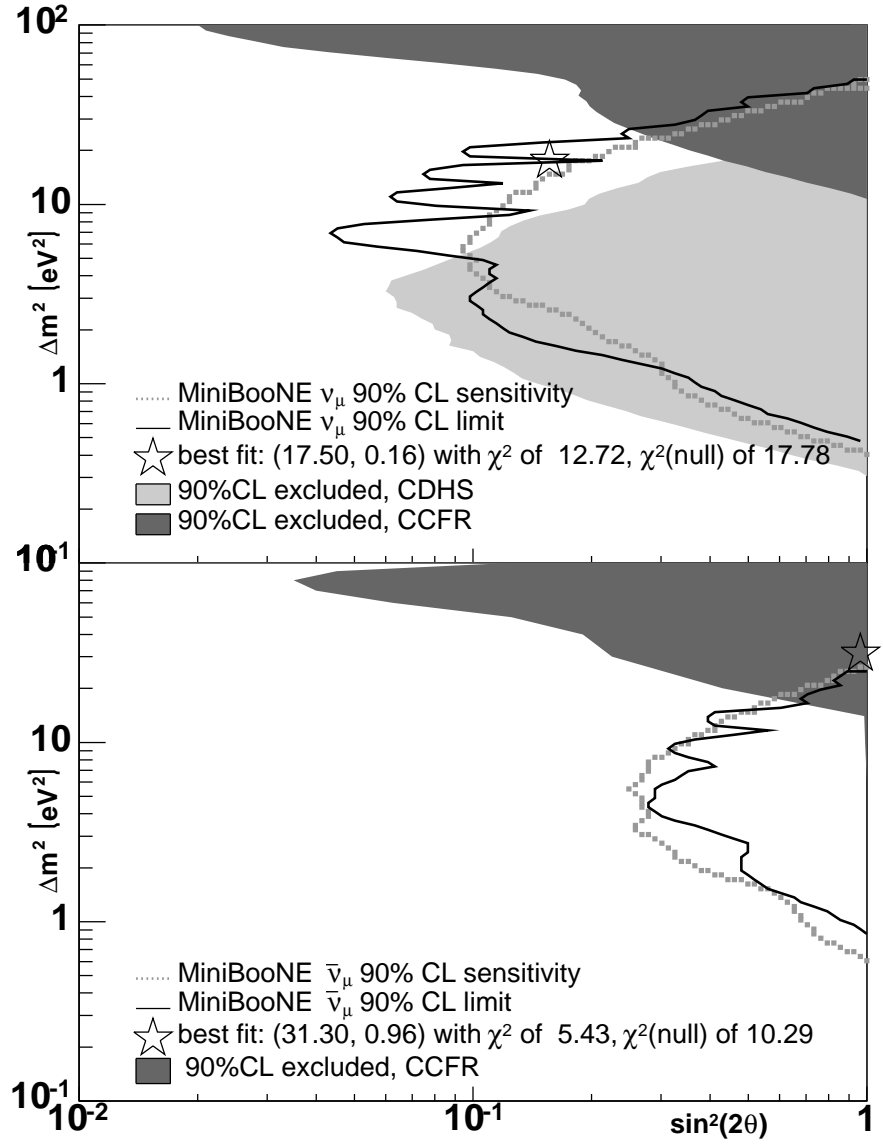}}
\caption{\label{disappearance} \em The sensitivities and
limits at 90\% CL for $\nu_\mu$ disappearance (top plot) and $\bar \nu_\mu$
disappearance (bottom plot).}
\end{figure}

MiniBooNE has also searched for $\nu_\mu$ and $\bar \nu_\mu$ disappearance 
\cite{mb_disappearance}. Fig. \ref{disappearance} shows the sensitivities and
limits at 90\% CL for $\nu_\mu$ disappearance (top plot) and $\bar \nu_\mu$
disappearance (bottom plot). The stars on the plots show the best fits in each
case: $\Delta m^2 = 17.50$ eV$^2$, $\sin^22\theta = 0.16$, and $\chi^2=12.72$/14 DF
for $\nu_\mu$ disappearance and $\Delta m^2 = 31.30$ eV$^2$, $\sin^22\theta = 0.96$, and 
$\chi^2=5.43$/14 DF for $\bar \nu_\mu$ disappearance. The $\chi^2$ values for
no disappearance oscillations 
are $\chi^2=17.78$/16 DF and $\chi^2=10.29$/16 DF, respectively.
Improved disappearance sensitivities are expected with the joint SciBooNE/MiniBooNE
analysis, which should be completed soon. Note, however, that the
joint SciBooNE/MiniBooNE analysis will not be nearly as powerful as the
joint BooNE/MiniBooNE analysis due to the small size of SciBooNE and 
the larger systematic errors from SciBooNE's different detector technology.

\section{BooNE}
The BooNE experiment involves building a second detector at a cost
of $\sim \$8$M along the BNB at FNAL at a closer distance of $\sim 200$ m.
With two detectors, many of the systematic errors will cancel,
as the neutrino flux varies as $1/r^2$ to good approximation, so that a 
ratio of events in the two detectors will provide a sensitive search
for $\nu_e$ and $\bar \nu_e$ appearance and $\nu_\mu$ and $\bar \nu_\mu$
disappearance. Furthermore, by
comparing the rates for a NC reaction, such as NC
$\pi^0$ scattering or NC elastic scattering,
a direct search for sterile neutrinos can be made.
An even cheaper option would be to move the MiniBooNE detector to a different
location at a cost of only $\sim \$4$M. If the MiniBooNE detector
were moved to a distance of 200 m from the neutrino source, then the event
rate would increase by a factor of $\sim 7$ due to the 
dependence of the neutrino flux on distance. An additional advantage of
moving MiniBooNE is that MicroBooNE could then move into
the original MiniBooNE enclosure and, therefore, 
save the expense of building a new
MicroBooNE enclosure. In either case,
after less than a year of running, the comparison
of the event rates at the two locations will determine whether the low-energy
excess observed by MiniBooNE was due to neutrino oscillations. In addition,
$\nu_\mu$ and $\bar \nu_\mu$ disappearance will be searched for with high
sensitivity in the $\Delta m^2 > 0.1$ eV$^2$ mass region, and LSND
antineutrino oscillations can be tested directly by searching for
$\bar \nu_e$ appearance. By comparing 
neutrino oscillations to antineutrino oscillations, BooNE will be able
to search for CP and CPT violation in the lepton sector at short baseline
($\Delta m^2 > 0.1$ eV$^2$). For the sensitivities discussed below, it is
assumed that the near detector will run for $\sim 1$E20 POT in both
neutrino mode and antineutrino mode.

\subsection{Fluxes and Event Rates}

This section gives a detailed comparison of the expected neutrino
fluxes at the near (200 meter) and far (541 meters) positions. In the
Booster neutrino beam (BNB), the primary beam is produced by the 8 GeV
Fermilab's rapid-cycling (15Hz) booster accelerator, which produces
1.6 $\mu s$ batches of protons each containing around
$4.5\times 10^{12}$ protons.

At that primary proton energy, there are only four significant species
of neutrinos: $\numu$ and $\numubar$ ($\sim$ 99.5\%), and a small
contamination ($\sim$ 0.5\%) of $\nue$ and $\nuebar$. There are two primary
parent components to the fluxes: neutrinos from charged pion decays
and neutrinos from kaon decays. The $K^+$ component dominates the
$\numu$ spectrum above neutrino energies of 2.5 GeV, where a clear break
is observed in the slope of the energy spectrum. The $\numubar$ spectra
are mainly from charged pion decay, and the $\nue$ and $\nuebar$
spectra are composed of two parts, muon decays and kaon decays.

The standard MiniBooNE Geant4 based beam simulation and decay program
packages were used to generate fluxes\cite{mb_flux}. Those
packages include the transport of muon polarization (neglecting $g-2$
precession effects) and appropriate form factors in leptonic kaon
decays. The primary production of pions by 8 GeV protons was measured
by the HARP experiment\cite{harp} and is used as input in the simulation, while
secondary interactions in the beam line are handled by standard Geant4
physics packages.

The fluxes shown here represent the spectrum of neutrinos that
intersect a sphere of radius 610.6 cm, positioned at either the near
or far location. The fluxes are ``unoscillated'' and therefore have
only $\numu$($\numubar$)and $\nue$($\nuebar$) components. No matter
effects in propagating the neutrinos to the detector are included, as
they are expected to be small in the standard, 3-generation, active
neutrino model (S$\nu$M).

\begin{figure}[h!]
%\begin{minipage}{14pc}
\centerline{\includegraphics[height=3.50in]{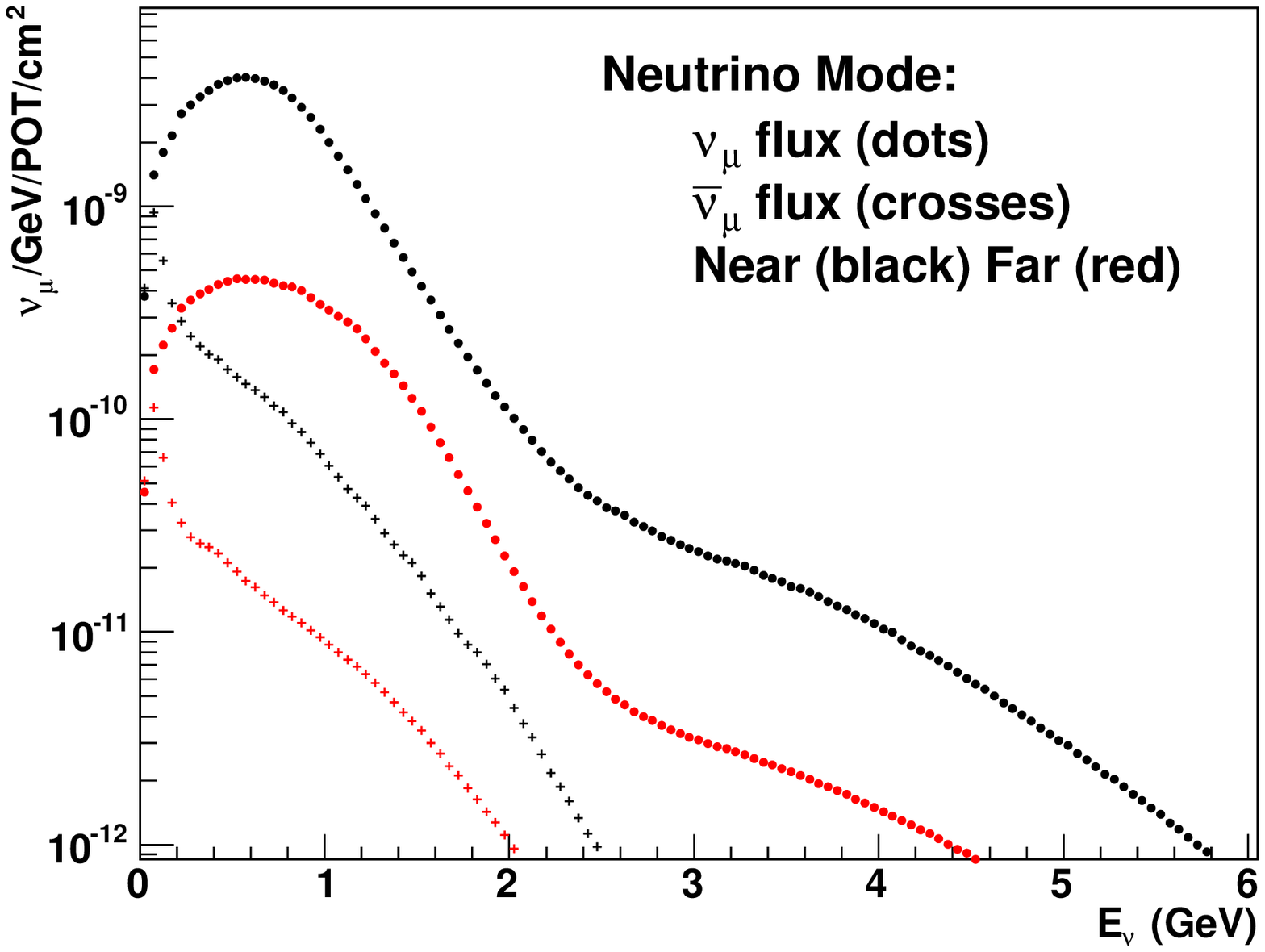}}
%\centerline{\psfig{file=geography.eps,width=18.cm}}
%\includegraphics[width=38pc]{geography.eps}
\caption{\label{numodenumu} \em The $\numu$ and $\numubar$ fluxes at both the near
 and far locations in neutrino mode.}
%\end{minipage}\hspace{2pc}%
%\begin{minipage}{14pc}
%\includegraphics[width=14pc]{name.eps}
%\caption{\label{label}Figure caption for second of two sided figures.}
%\end{minipage}
\end{figure}

\begin{figure}[h!]
%\begin{minipage}{14pc}
\centerline{\includegraphics[height=3.50in]{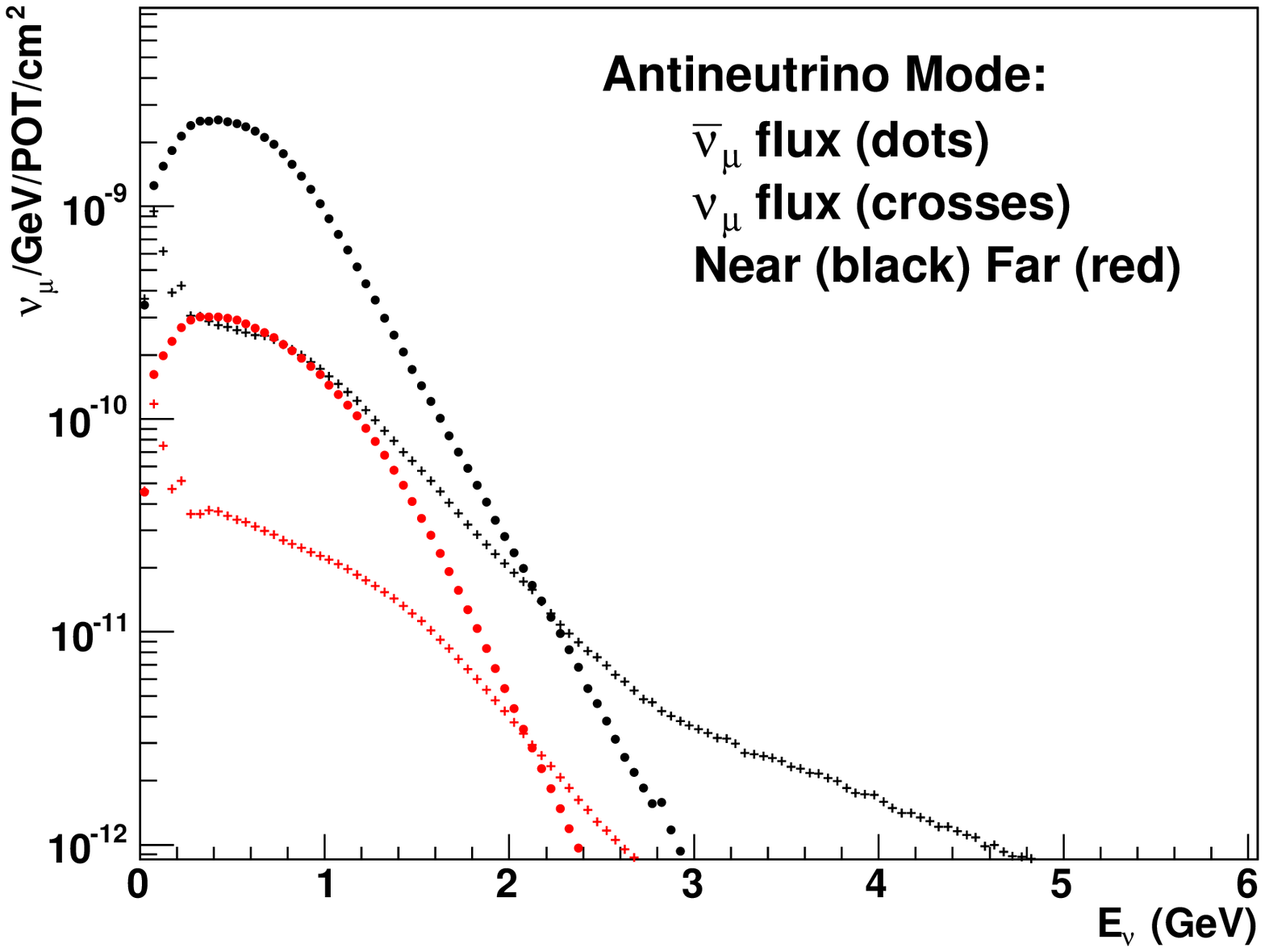}}
%\centerline{\psfig{file=geography.eps,width=18.cm}}
%\includegraphics[width=38pc]{geography.eps}
\caption{\label{nubarmodenumu} \em The $\numu$ and $\numubar$ fluxes at both the near
 and far locations in antineutrino mode.}
%\end{minipage}\hspace{2pc}%
%\begin{minipage}{14pc}
%\includegraphics[width=14pc]{name.eps}
%\caption{\label{label}Figure caption for second of two sided figures.}
%\end{minipage}
\end{figure}

\begin{figure}[h!]
%\begin{minipage}{14pc}
\centerline{\includegraphics[height=3.50in]{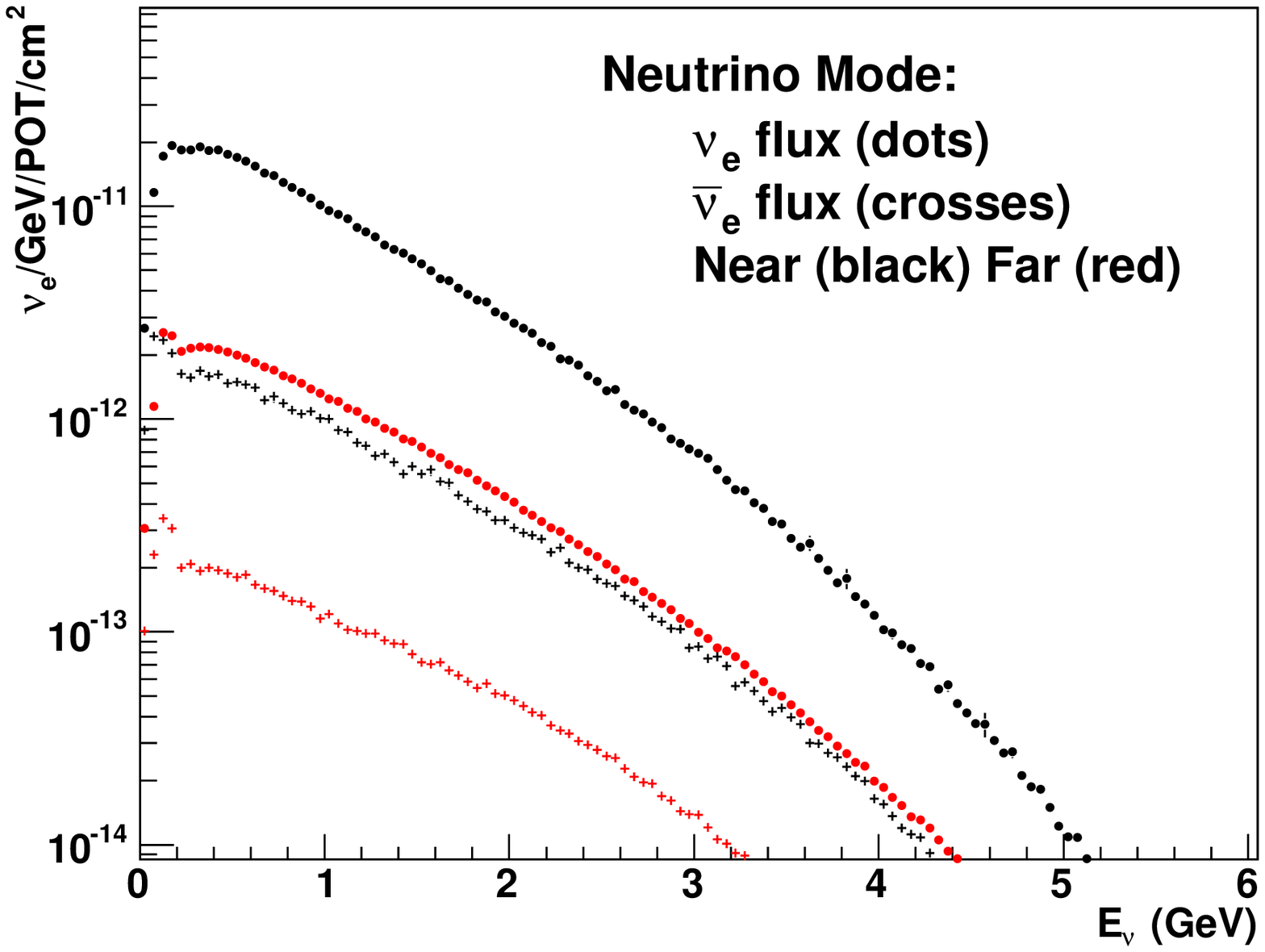}}
%\centerline{\psfig{file=geography.eps,width=18.cm}}
%\includegraphics[width=38pc]{geography.eps}
\caption{\label{numodenue} \em The $\nue$ and $\nuebar$ fluxes at both the near
 and far locations in neutrino mode.}
%\end{minipage}\hspace{2pc}%
%\begin{minipage}{14pc}
%\includegraphics[width=14pc]{name.eps}
%\caption{\label{label}Figure caption for second of two sided figures.}
%\end{minipage}
\end{figure}

\begin{figure}[h!]
%\begin{minipage}{14pc}
\centerline{\includegraphics[height=3.50in]{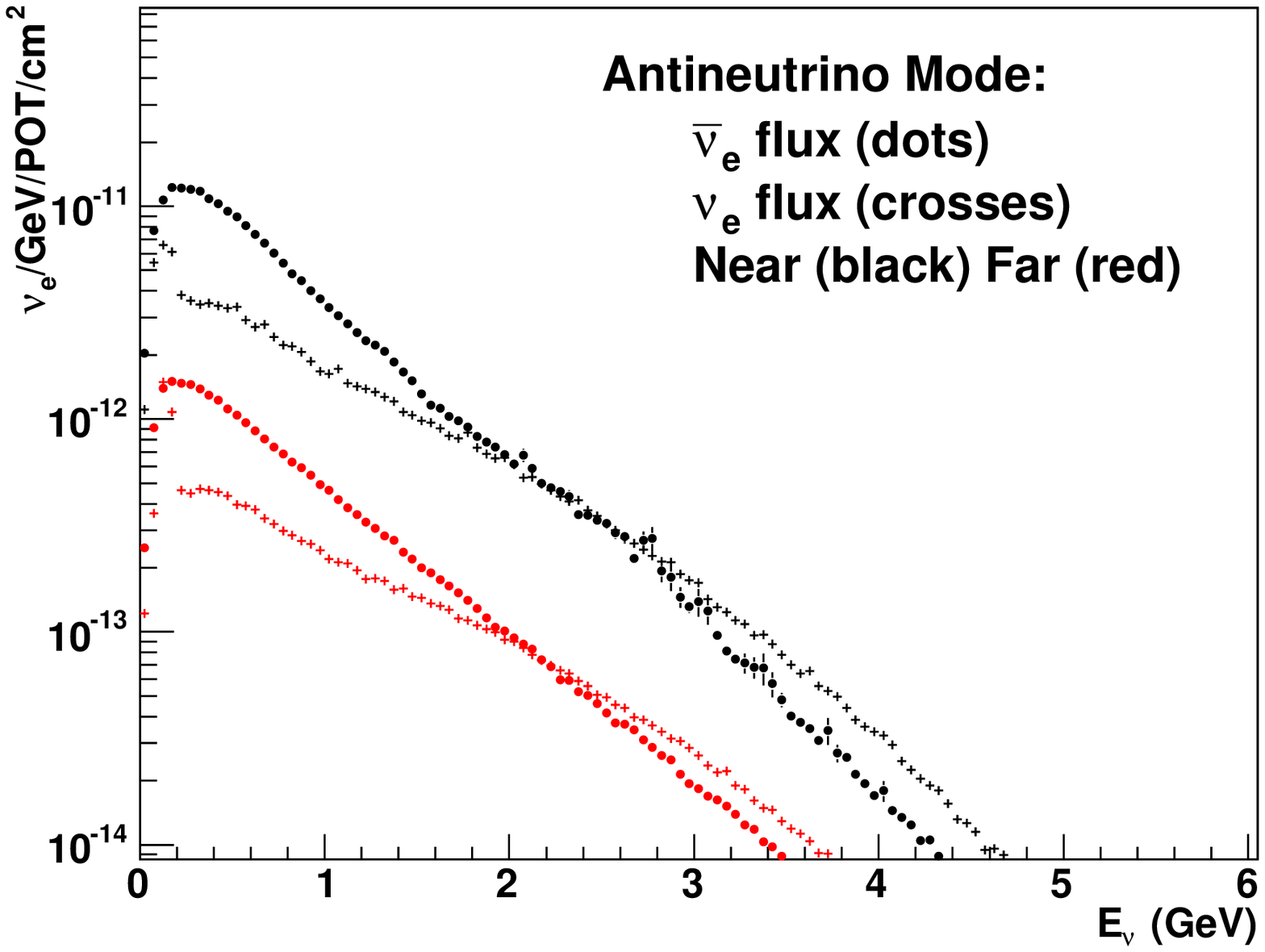}}
%\centerline{\psfig{file=geography.eps,width=18.cm}}
%\includegraphics[width=38pc]{geography.eps}
\caption{\label{nubarmodenue} \em The $\nue$ and $\nuebar$ fluxes at both the near
 and far locations in antineutrino mode.}
%\end{minipage}\hspace{2pc}%
%\begin{minipage}{14pc}
%\includegraphics[width=14pc]{name.eps}
%\caption{\label{label}Figure caption for second of two sided figures.}
%\end{minipage}
\end{figure}

Figs. \ref{numodenumu}, \ref{nubarmodenumu}, \ref{numodenue}, and
\ref{nubarmodenue} show the fluxes for the four neutrino species at
the near and far locations, for both neutrino mode and antineutrino
mode. Table \ref{tab:fluxes} gives the same fluxes, integrated over
neutrino energy, while Table \ref{tab:energies} gives the average
neutrino energy in each case.

In neutrino mode, the $\numu$ flux near/far ratio is 7.5. Most of
the near far ratios are between 7.0 and 8.0. Another characteristic of
the near/far flux comparisions is that the average energy of the
neutrinos in the near position is between 5 and 10 percent less than
the corresponding average energy in the far position. This lower energy is
expected since the near detector has a larger angular acceptance with
respect to the neutrino target.

\begin{figure}[h!]
%\begin{minipage}{14pc}
\centerline{\includegraphics[height=3.50in]{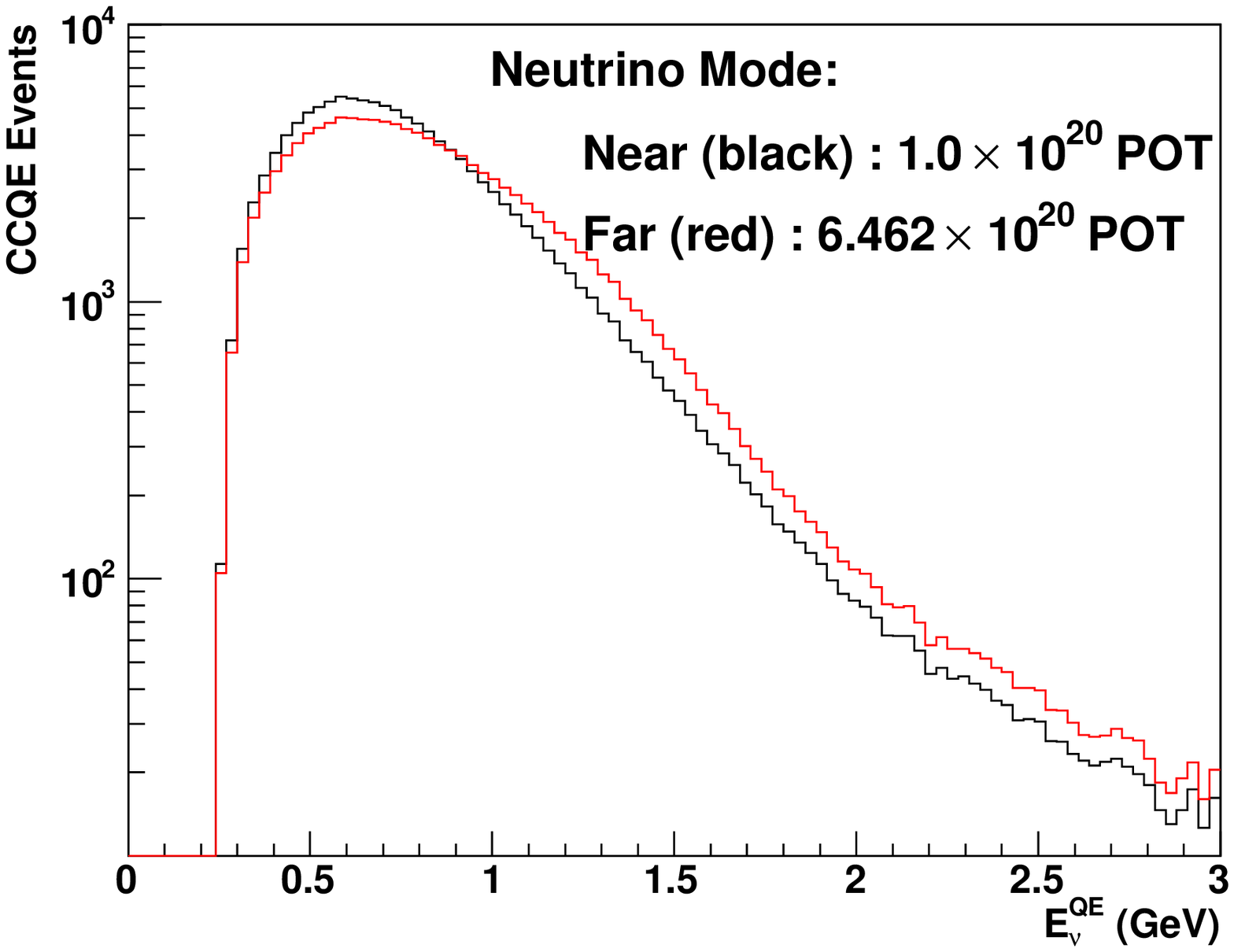}}
%\centerline{\psfig{file=geography.eps,width=18.cm}}
%\includegraphics[width=38pc]{geography.eps}
\caption{\label{numodeccqe} \em The distribution of energies for reconstructed
$\nu_\mu$ CCQE events, at the near location with
$1.0\times 10^{20}$ POT and at the far location with the current
$6.462\times 10^{20}$ POT.}
%\end{minipage}\hspace{2pc}%
%\begin{minipage}{14pc}
%\includegraphics[width=14pc]{name.eps}
%\caption{\label{label}Figure caption for second of two sided figures.}
%\end{minipage}
\end{figure}

Fig. \ref{numodeccqe} shows the energy distribution for $\nu_\mu$ CCQE
events at the near ($1.0\times 10^{20}$ POT) and
far ($6.462\times 10^{20}$ POT) locations for neutrino mode. The
spectral differences are again due to the larger angular acceptance of
the near detector. That larger decay angle translates to lower neutrino
energies in the near detector, typically $\sim 10\%$ lower in the 200/541
meter comparison. This extrapolation is relatively straight forward
as the angular divergence of the daughter neutrinos in the 
decays is much larger than the angular divergence of the decaying
mesons themselves. For example, even at 3 GeV, daughter neutrinos from
pion and kaon decays will have opening angles of $\sim$ 50 mrad and $\sim$ 150
mrad, respectively, while the allowed angular divergence of the beam tunnel is
only $\sim$ 20 mrad.

Because of the nearly complete overlap in decay particle phase space
that contributes to neutrinos in the near and far positions, we expect
that uncertainties in the flux prediction will largely cancel when
comparing the two event rates from the near and far positions. As
systematic errors introduced by uncertainties in the detector efficiency and
neutrino cross section will also largely cancel, the comparison of the
two positions will allow a much-needed, accurate measurement of
non-S$\nu$M neutrino oscillation effects in the $\Delta m^2$ range of
0.1-10 $eV^2$.

\begin{table}[h!]
%\begin{table}
\begin{center}
\caption{\label{tab:fluxes} \em Integrated fluxes per POT for the various species of neutrinos at the
near and far positions, for both neutrino mode and antineutrino mode.}
%\begin{ruledtabular}
\vspace{0.1in}
\begin{tabular}{|c|c|c|c|c|}

\multicolumn{5}{c}{Fluxes $\nu/(cm^2\ POT)$}  \\
\hline
              & \multicolumn{2}{|c|}{$\nu$ mode} & \multicolumn{2}{|c|}{$\bar\nu$ mode}  \\
\hline
$\nu$ species & Near      & Far      & Near   & Far  \\
\hline
$\numu$    & $7.49\times 10^{-8}$  & $1.03\times 10^{-8}$  & $8.12\times 10^{-9}$ & $1.08\times 10^{-9}$ \\
$\numubar$ & $5.20\times 10^{-9}$  & $6.52\times 10^{-10}$ & $4.30\times 10^{-8}$ & $5.77\times 10^{-9}$ \\
$\nue$     & $4.50\times 10^{-10}$ & $5.74\times 10^{-11}$ & $9.5\times 10^{-11}$ & $1.34\times 10^{-11}$ \\
$\nuebar$  & $4.61\times 10^{-11}$ & $6.00\times 10^{-12}$ & $2.00\times 10^{-10}$ & $2.53\times 10^{-11}$ \\
\hline
\end{tabular}
\end{center}
\end{table}
%nue flux near   (nu mode)    : 4.50486e-10
%nue flux far    (nu mode)    : 5.73831e-11
%nuebar flux near(nu mode)    : 4.61474e-11
%nuebar flux far (nu mode)    : 5.99724e-12
%nue flux near (nubar mode)   : 9.52165e-11
%nue flux far   (nubar mode)  : 1.33577e-11
%nuebar flux near (nubar mode): 2.00065e-10
%nuebar flux far (nubar mode) : 2.53112e-11

%numu flux near   (nu mode)    : 7.48721e-08
%numu flux far    (nu mode)    : 1.0303e-08
%numubar flux near(nu mode)    : 5.19786e-09
%numubar flux far (nu mode)    : 6.5195e-10
%numu flux near (nubar mode)   : 8.11898e-09
%numu flux far   (nubar mode)  : 1.08435e-09
%numubar flux near (nubar mode): 4.29762e-08
%numubar flux far (nubar mode) : 5.76732e-09

\begin{table}[h!]
\begin{center}
\caption{\label{tab:energies} \em Average neutrino energies for the various species of neutrinos at the
near and far positions, for both neutrino mode and antineutrino mode.}
\vspace{0.1in}
\begin{tabular}{|c|c|c|c|c|}

\multicolumn{5}{c}{Average $\nu$ energies (MeV)}  \\
\hline
              & \multicolumn{2}{|c|}{$\nu$ mode} & \multicolumn{2}{|c|}{$\bar\nu$ mode}  \\
\hline
$\nu$ species & Near      & Far      & Near   & Far  \\
\hline
$\numu$ & $721$ & $807$ & $631$ & $703$ \\
$\numubar$ & $412$ & $461$ & $593$ & $649$ \\
$\nue$ & $903$ & $957$ & $856$ & $874$ \\
$\nuebar$ & $917$ & $971$ & $677$ & $716$ \\
\hline
\end{tabular}
\end{center}
\end{table}

%numu flux near   (nu mode)    : 0.721564
%numu flux far    (nu mode)    : 0.807069
%numubar flux near(nu mode)    : 0.412504
%numubar flux far (nu mode)    : 0.461292
%numu flux near (nubar mode)   : 0.631083
%numu flux far   (nubar mode)  : 0.70334
%numubar flux near (nubar mode): 0.593314
%numubar flux far (nubar mode) : 0.64885

%nue flux near   (nu mode)    : 0.903533
%nue flux far    (nu mode)    : 0.957536
%nuebar flux near(nu mode)    : 0.917201
%nuebar flux far (nu mode)    : 0.971247
%nue flux near (nubar mode)   : 0.856822
%nue flux far   (nubar mode)  : 0.874376
%nuebar flux near (nubar mode): 0.676533
%nuebar flux far (nubar mode) : 0.715944

\subsection{Possible Scenarios for a Near Detector}

The MiniBooNE detector has operated at a location of 541 meters from the
Booster Neutrino Beam (BNB) target since September 1, 2002. The primary
purpose of the experiment was to search for the transmutation, or
oscillation, of muon neutrinos into electron neutrinos as they travel
the $\sim$ 525 meters to the detector. The BNB was designed to produce
a nearly pure beam of $\numu$, which provides an ideal setting to look
for excess $\nue$ events. While the most sensitive neutrino
oscillation experiments are two detector systems, which afford a
comparison of similar detectors at two different distances, MiniBooNE
was built as a single detector system in order to reduce costs. It was
felt that the systematic error incurred by not building a second
detector could be overcome by using internal measurements in the
single detector. As $\numu$ were not expected to oscillate
significantly, it was planned to use their rate as a normalization for
$\nue$ interactions, thus constraining backgrounds to $\nue$ events
from oscillations.

The MiniBooNE proposal foresaw that a second detector at a different
distance would be required to ascertain the nature of the signal, if a
significant signal were observed in the single-detector setup.  A
second detector, located at a different distance from the BNB target
could potentially remove the large systematic errors that would
complicate the interpretation of the MiniBooNE data.

There are several possible routes to a more precise measurement which
need to be considered: a new detector could be constructed at a near
position; similarly, a new detector could be constructed at a far
position; and, it was recently realized that the MiniBooNE detector
could be relocated to a new position at either a near or a far
location. Each of those possibilities has advantages and
disadvantages.

\subsubsection{Near or Far?}
The choice between constructing a near detector at $\sim$ 200 meters
and a far detector at $\sim$ 1000 meters can be made based on
expediency. For $\Delta m^2 < 2$ eV$^2$, 
a near detector will not see a large signal directly but
can be used to accurately measure the expected backgrounds to any
possible oscillation signal in the far detector. Those backgrounds, in 
both appearance and disappearance measurements, can be measured at
$\sim$ 7-8 times the rate that the MiniBooNE detector accumulated
data. Thus a sample of neutrinos with statistics equivalent to
MiniBooNE's existing data set of $6.462\times 10^{20}$ POT will only
require $\sim\ 1.0\times 10^{20}$ POT and yield an $\sim$ 5$\sigma$
result. An identical far detector, on the other hand, would also yield
an $\sim$ 5$\sigma$ result with $\sim\ 1.0\times 10^{20}$ POT, however
the signal would only be $\sim$ 20 events on top of a background of
$\sim$ 16 events. An unsatisfying, ambiguous result could occur with
such low statistics.

\subsubsection{Moving MiniBooNE}

\begin{figure}[ht]
%\begin{minipage}{14pc}
\centerline{\includegraphics[height=3.50in]{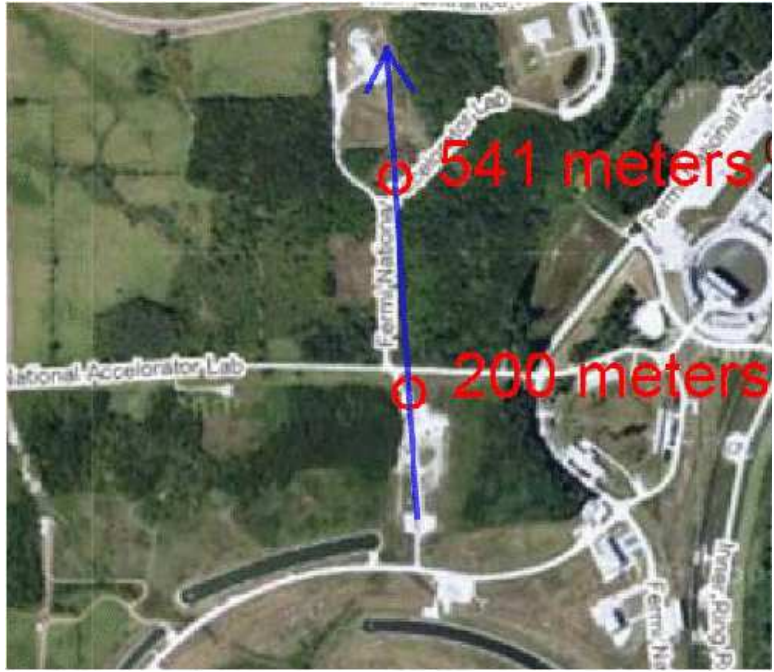}}
%\centerline{\psfig{file=geography.eps,width=18.cm}}
%\includegraphics[width=38pc]{geography.eps}
\caption{\label{fnal-arial} \em An arial view of Fermilab showing
a possible location of a near detector at $\sim$ 200 meters. The blue
arrow indicates the direction of the booster neutrino beam.}
%\end{minipage}\hspace{2pc}%
%\begin{minipage}{14pc}
%\includegraphics[width=14pc]{name.eps}
%\caption{\label{label}Figure caption for second of two sided figures.}
%\end{minipage}
\end{figure}

Relocating the MiniBooNE detector to a near position $\sim$ 200 meters
from the target, shown in Fig. \ref{fnal-arial}, will likely be the
least costly option. There are a number of potential advantages to this
approach. The detector is already built so that the cost of
constructing a new detector is avoided. The neutrino flux at 200
meters will be $\sim$ 7-8 times larger, so that the time needed to
accumulate a data sample equivalent to the existing sample will be
less than one year.

The comparison of measurements taken by the same detector operated at
two locations will be free of systematic errors associated with
neutrino-nucleus cross sections and detector response. The comparison
should clearly reveal the nature of the MiniBooNE excess. The 541/200
meter data comparison will also allow MiniBooNE to measure
$\nu_\mu$ and $\bar \nu_\mu$ disappearance at the few percent level.

The detector relocation could take place in two ways:
transporting the existing detector and 
electronics to the new location, or building a new detector at the 
new location using parts from the existing detector. In both cases
the MiniBooNE dectector would be drained of oil and the oil stored in
rail cars or a separate oil storage tank. With the mineral oil
removed, the detector weighs only $\sim$ 80 tons, and could be
transported whole to the near site. Mobile cranes with 80 ton lfting
capacities are readily available commercially at a reasonable cost.

\subsubsection{A New Detector}
An alternative to relocating the existing MiniBooNE detector 
and its electronics is to construct a new sperical tank at the near 
position and remove PMTs by hand from the old tank.
The new tank
would be constructed by repeating the constuction effort made for
MiniBooNE. In either case, one would re-use the existing, albeit old,
LSND electronics currently used by MiniBooNE. The cost of either case
is estimated to be $\sim$ \$4M.

The most desirable, and most costly, option is to construct an
entirely new detector at the near location. This would require more
time because new electronics would have to be developed, a new oil
mineral supplier found, and new phototubes purchased. The lead times on
those items would be about two years. The cost for that effort is
estimated to be $\sim$ twice that of moving the existing detector,
$\sim$ \$8M.

It is not yet understood how the systematic error in detector response
will translate between the old MiniBooNE detector and the newly
constructed detector, since it will have different oil, PMTs, and
electronics. Nevertheless, choosing to construct an entirely new
detector would allow for simultaneous operation of both the near and
far detector and eliminate any fear, unfounded or not, that the
neutrino beam had changed in character.

Ideally, the near BooNE detector would have the same dimensions as the
MiniBooNE detector in order to reduce systematic uncertainties. However,
another possibility would be to build a smaller detector ($\sim 8$ m
diameter) at a lower cost ($\sim \$4$M) if systematic errors were estimated
to be sufficiently small.

\subsection{Testing the Low-Energy Excess}

BooNE will be able to determine
whether the low-energy excess is due to neutrino oscillations and
will be able to test various hypotheses by comparing the 
low-energy excesses in neutrino and antineutrino modes.
If the low-energy excess is due to
background, then the near detector will observe the same relative excess
as the far detector. If the excess is due to neutrino
oscillations at low $\Delta m^2$, then no low-energy excess will be observed
in the near detector and the current low-energy excess in the far detector,
assuming a 2.5\% systematic error, will 
equal to $128.8 \pm 20.4 \pm 10.4$ events ($5.6 \sigma$).
For testing various hypotheses, 
Table \ref{implications2} shows the expected excess
of low-energy events ($200< E_\nu <475$ MeV) in antineutrino mode for 1E21 POT, assuming a
2.5\% systematic error and assuming that the neutrino excess is correct. 
Also shown is the significance of the excesses.
By comparing the measured excess in the antineutrino data
to the expected excesses, the different hypotheses can be shown to be
either consistent with data or ruled out. 

\begin{table}
%\begin{table}[h]
\begin{center}
\caption{\label{implications2} \em The expected excess
of low-energy events ($200< E_\nu <475$ MeV)
in antineutrino mode (1E21 POT) for
various hypotheses, assuming a
2.5\% systematic error and assuming that the neutrino 
low-energy excess is correct. 
Also shown is the significance of the excesses.}
%%\begin{ruledtabular}
\vspace{0.1in}
\begin{tabular}{|c|c|c|}
\hline
Hypothesis& Expected Excess of $\bar \nu$ Events&Significance \\
\hline
Same $\sigma$&$111.6 \pm 13.5 \pm 4.6$&$7.8\sigma$ \\
$\pi^0$ Scaled&$58.2 \pm 13.5 \pm 4.6$&$4.1\sigma$ \\
POT Scaled&$202.6 \pm 13.5 \pm 4.6$&$14.2\sigma$ \\
BKGD Scaled&$62.8 \pm 13.5 \pm 4.6$&$4.4\sigma$ \\
CC Scaled&$61.2 \pm 13.5 \pm 4.6$&$4.3\sigma$ \\
Kaon Scaled&$119.2 \pm 13.5 \pm 4.6$&$8.4\sigma$ \\
Neutrino Scaled&$20.2 \pm 13.5 \pm 4.6$&$1.4\sigma$ \\
\hline
\end{tabular}
%\end{center}
%%\end{ruledtabular}
\end{center}
\end{table}

\subsection{$\nu_e$ and $\bar \nu_e$ Appearance}

\begin{figure}
\centerline{\includegraphics[height=7.0in]{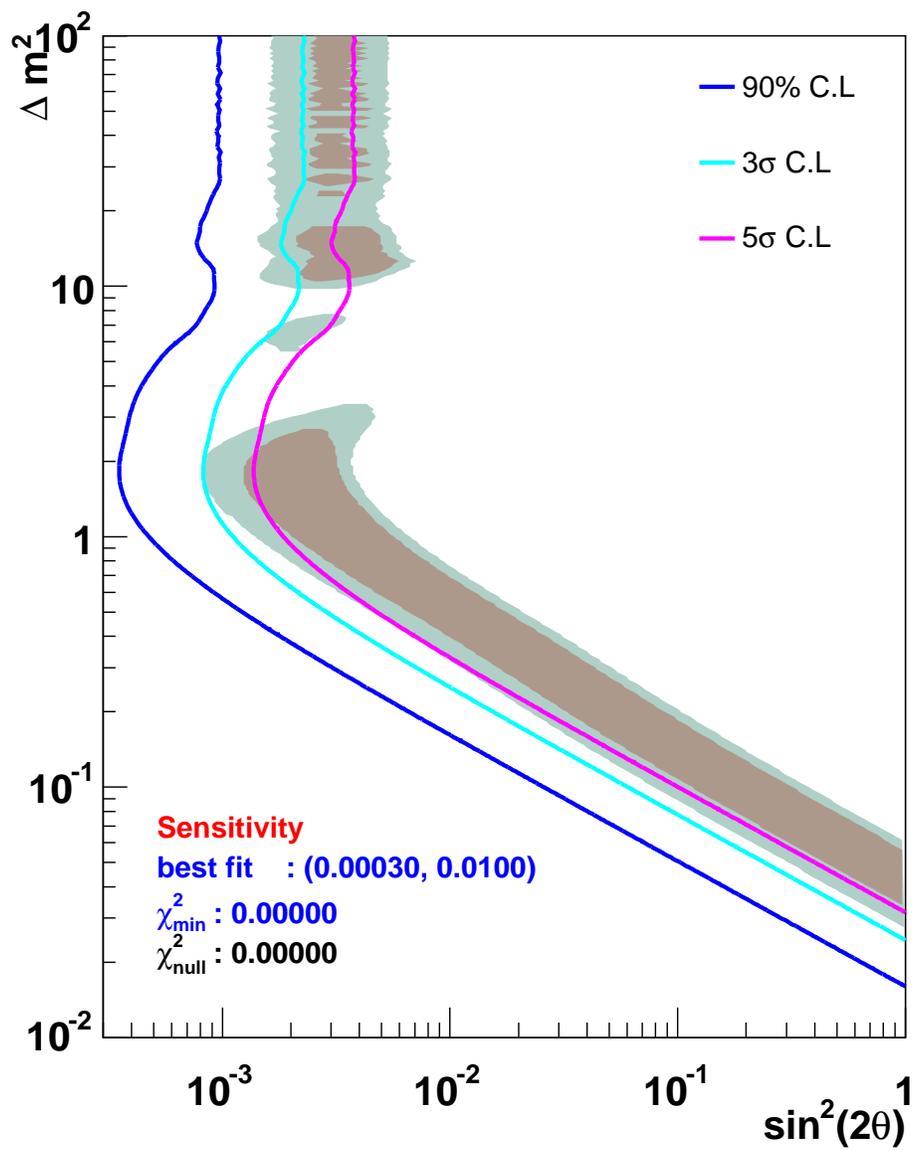}}
\caption{\label{nue_app_limit} \em The estimated BooNE sensitivity for 
$\nu_e$ appearance.}
\end{figure}

\begin{figure}
\centerline{\includegraphics[height=7.0in]{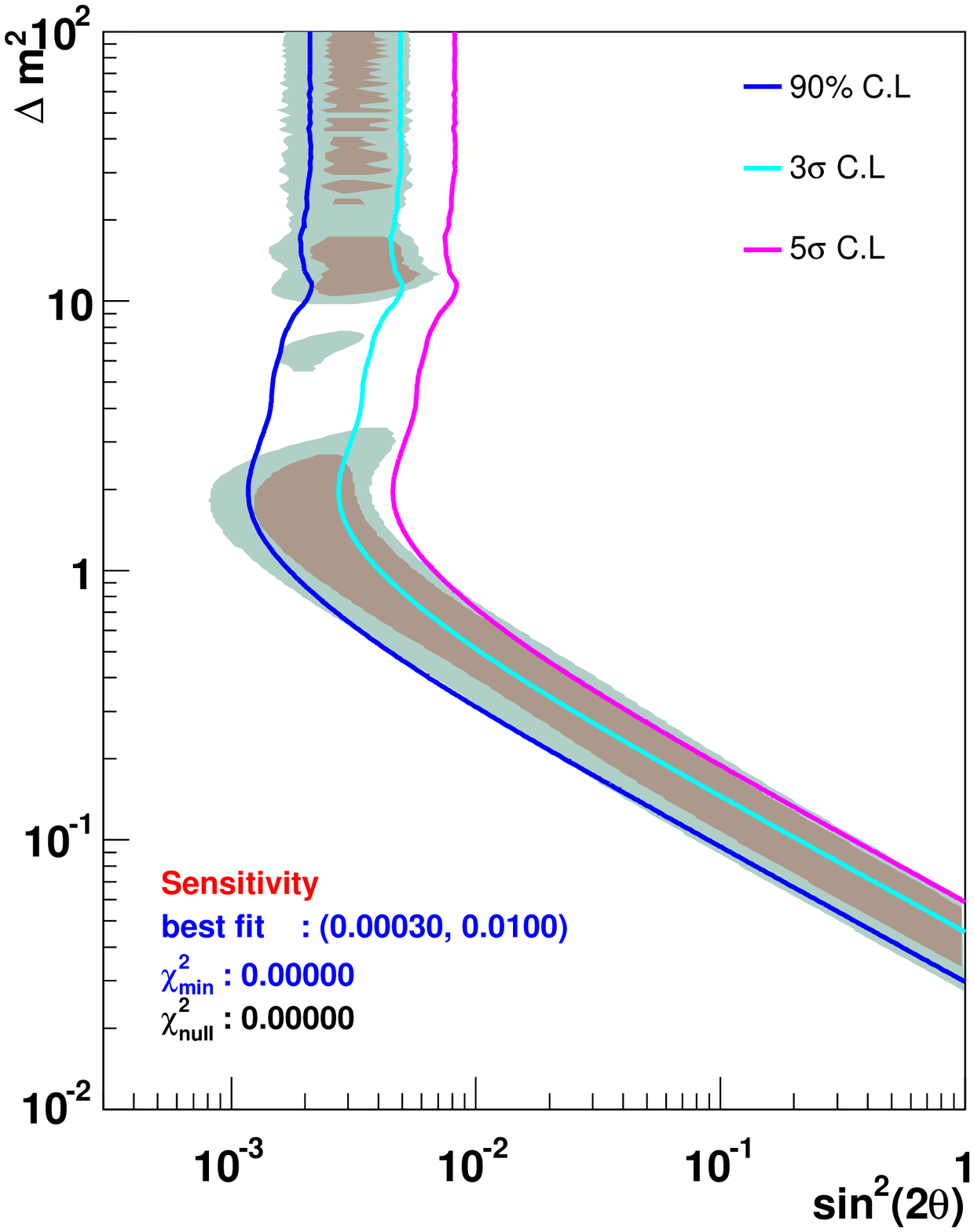}}
\caption{\label{nuebar_app_limit} \em The estimated BooNE sensitivity for 
$\bar \nu_e$ appearance.}
\end{figure}

The sensitivities for $\nu_e$ and $\bar \nu_e$ appearance
were obtained by using the
MiniBooNE Monte Carlo simulation, assuming statistical errors with the
expected MiniBooNE statistics (6.5E20 POT in neutrino mode and 1E21 POT
in antineutrino mode) and
a full error matrix with correlated and uncorrelated systematic errors.
Also, we assume 2E20 POT
in the near detector, equally divided between neutrino and antineutrino 
modes. Fig. \ref{nue_app_limit} shows
the estimated sensitivity for $\nu_e$ appearance for $E_\nu > 200$ MeV. 
Although simple two-neutrino 
oscillations have already been ruled out as an explanation of the LSND signal,
it is interesting that the full LSND region is covered at the approximately
5$\sigma$ level. Therefore, we would be able to determine whether or not the 
MiniBooNE low-energy excess is due to a more complicated oscillation 
mechanism at the $\sim 1$ eV$^2$ scale. Fig. \ref{nuebar_app_limit} shows
the estimated sensitivity for $\bar \nu_e$ appearance, where we assume 
that the error matrix is the same as for neutrinos. The sensitivity is
worse than the $\nu_e$ appearance sensitivity due to the lower statistics and
higher wrong-sign background in antineutrino mode; however, BooNE will still
be able to cover the full LSND region at 90\% CL and provide a direct test of LSND
antineutrino oscillations.

\subsection{$\nu_\mu$ and $\bar \nu_\mu$ Disappearance}

\begin{figure}
\centerline{\includegraphics[height=3.0in]{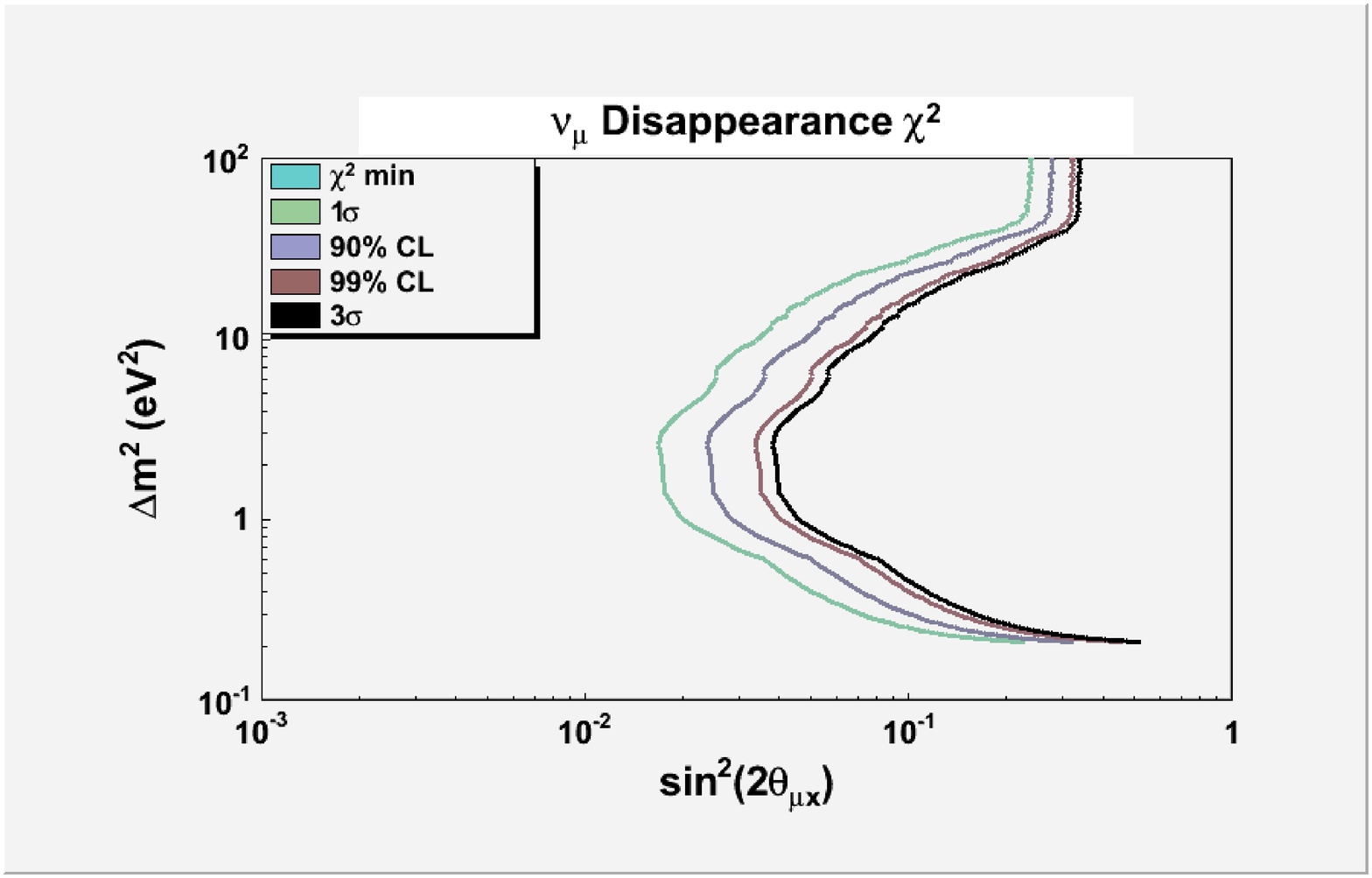}}
\caption{\label{numu_disapp_limit} \em The estimated BooNE sensitivity for 
$\nu_\mu$ disappearance.}
\end{figure}

\begin{figure}
\centerline{\includegraphics[height=3.0in]{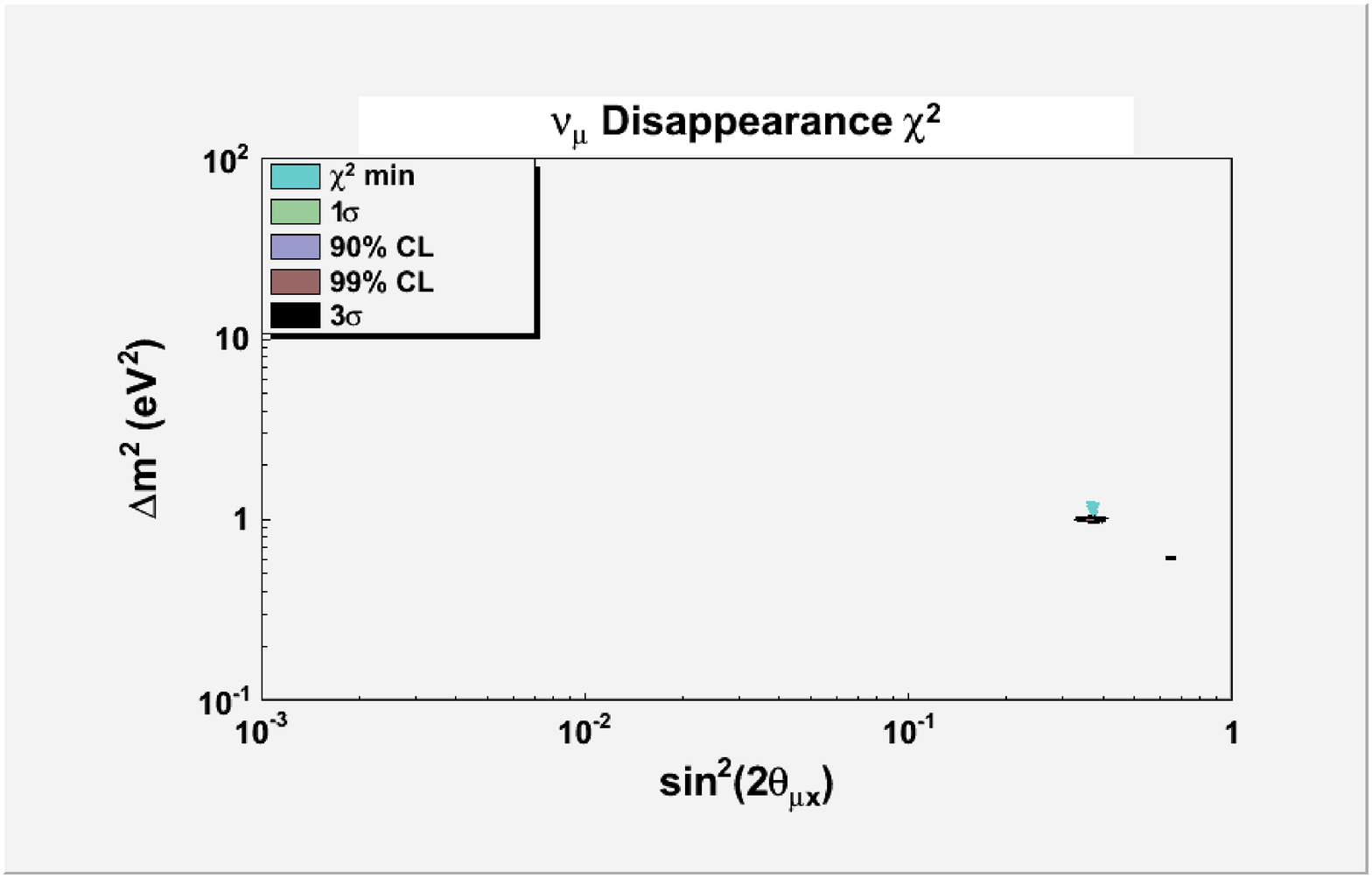}}
\caption{\label{numu_disapp_sig} \em The BooNE allowed region for 
$\nu_\mu$ disappearance at the global antineutrino best-fit
point $\Delta m_{14}^2 = 0.915$ eV$^2$ and
$\sin^22\theta_{\mu\mu} = 0.35$.}
\end{figure}

\begin{figure}
\centerline{\includegraphics[height=3.0in]{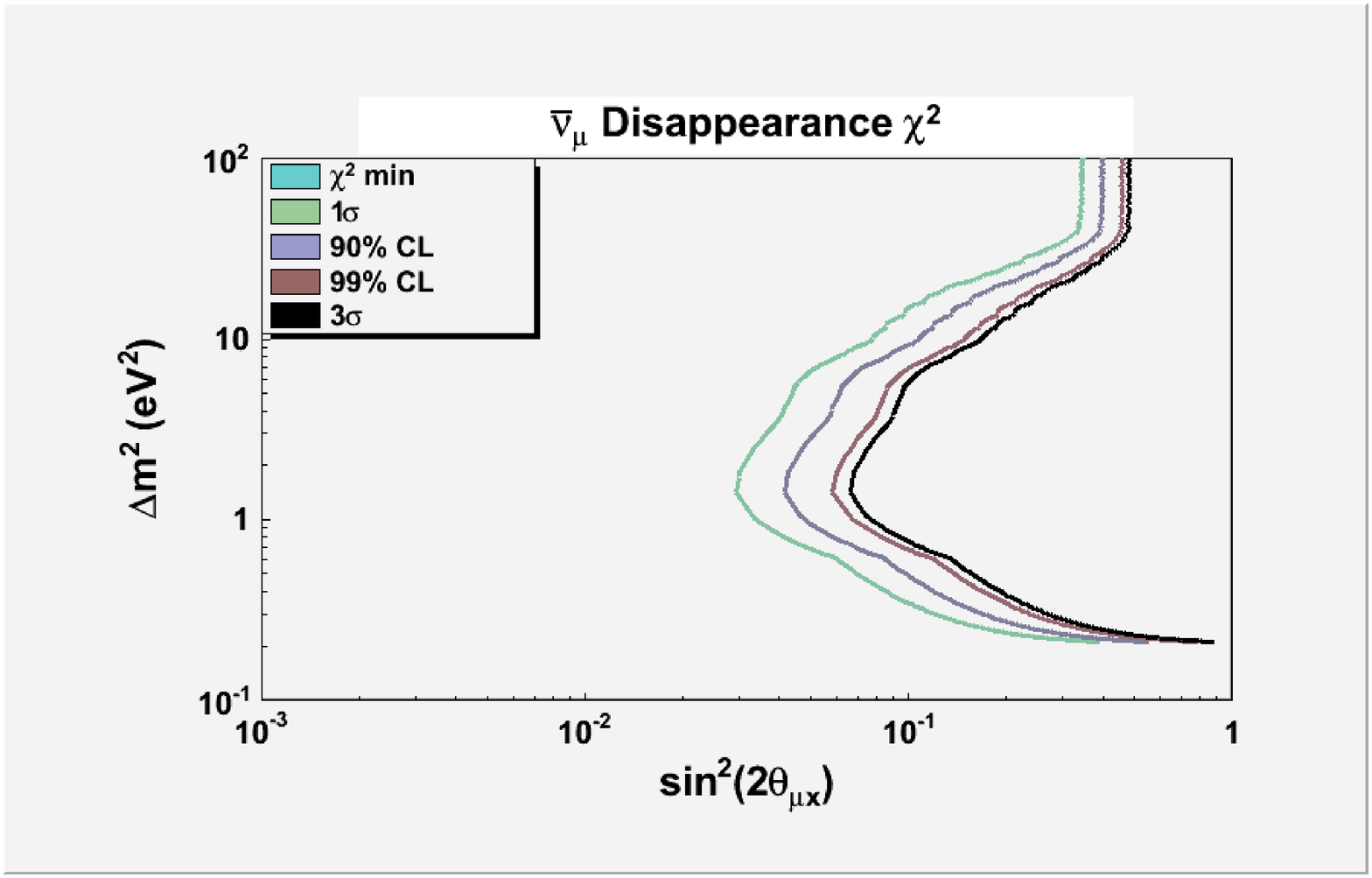}}
\caption{\label{antinumu_disapp_limit} \em The estimated BooNE sensitivity for
$\bar \nu_\mu$ disappearance.}
\end{figure}

\begin{figure}
\centerline{\includegraphics[height=3.0in]{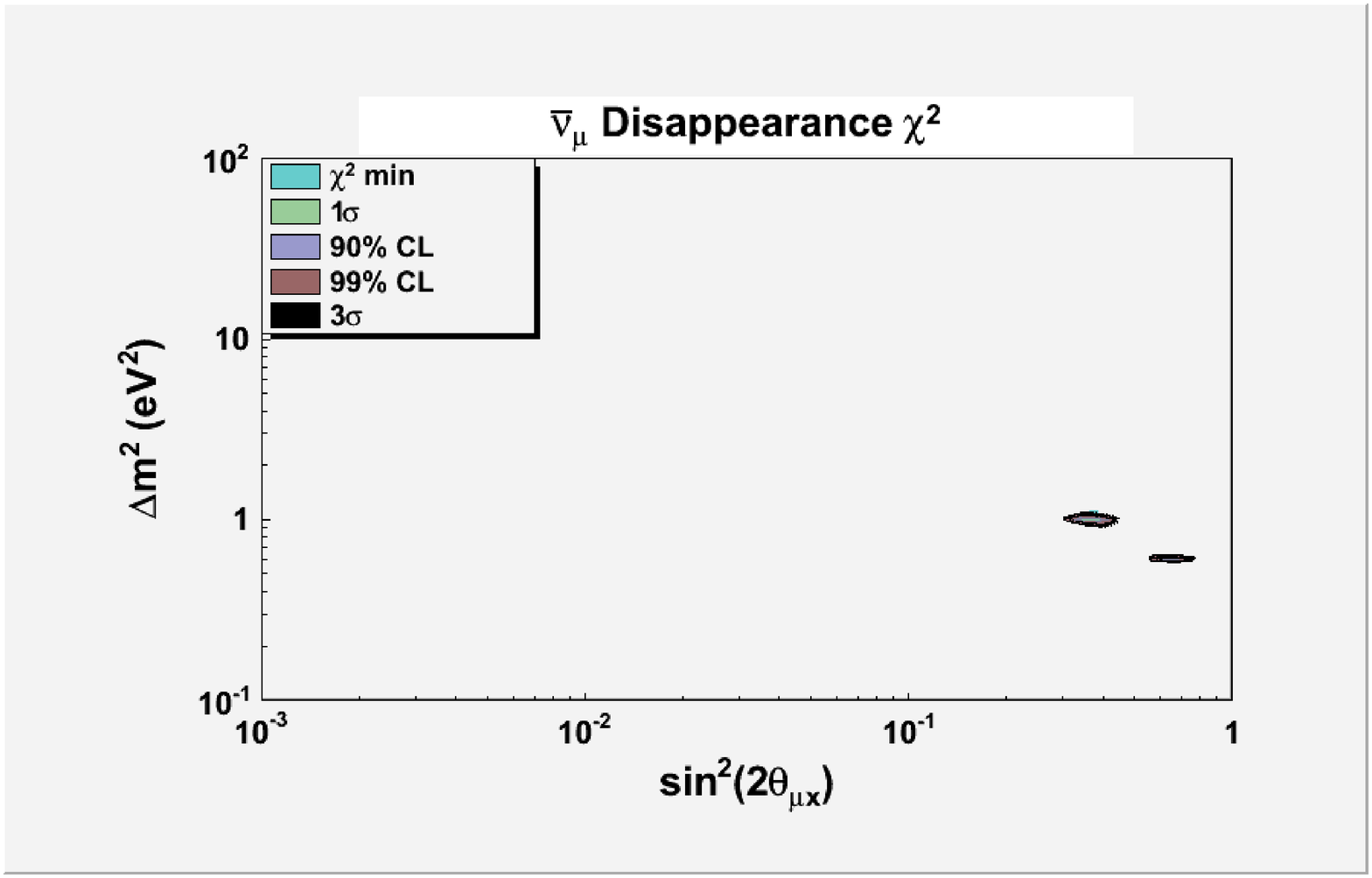}}
\caption{\label{antinumu_disapp_sig} \em The BooNE allowed region for
$\bar \nu_\mu$ disappearance at the global antineutrino best-fit
point $\Delta m_{14}^2 = 0.915$ eV$^2$ and
$\sin^22\theta_{\mu\mu} = 0.35$.}
\end{figure}

The sensitivities for $\nu_\mu$ and $\bar \nu_\mu$ disappearance 
were obtained by using the
MiniBooNE Monte Carlo simulation, assuming statistical errors with the
expected MiniBooNE statistics (6.5E20 POT in neutrino mode and 1E21 POT
in antineutrino mode) and a full error matrix with
correlated and uncorrelated systematic errors.
Also, we assume 2E20 POT
in the near detector, equally divided between neutrino and antineutrino
modes. Fig. \ref{numu_disapp_limit} shows
the estimated sensitivity for $\nu_\mu$ disappearance for $E_\nu >200$ MeV. 
A sensitivity of 
$\sim 3\%$ at 90\% CL is obtained for $\Delta m^2 \sim 1$ eV$^2$.
In order to see how a signal would appear, Fig. \ref{numu_disapp_sig}
shows the allowed region for $\nu_\mu$ disappearance at the global antineutrino
best-fit
point from reference \cite{georgia}: $\Delta m_{14}^2 = 0.915$ eV$^2$ and 
$\sin^22\theta_{\mu\mu} = 0.35$. Figs. \ref{antinumu_disapp_limit} and 
\ref{antinumu_disapp_sig} show the correponding limits and allowed regions for
$\bar \nu_\mu$ disappearance, assuming no $\nu_\mu$ disappearance and
the same error matrix as for neutrinos. 
The $\bar \nu_\mu$ sensitivity is slightly
worse than the $\nu_\mu$ sensitivity due to the lower antineutrino statistics and the
$\sim 1/3$ wrong-sign $\nu_\mu$ component in
antineutrino mode. A difference between $\nu_\mu$ and $\bar \nu_\mu$ disappearance
would be evidence for CPT violation or effective CPT violation \cite{gabriela,paes}.

\subsection{Sterile Neutrino Search}

If $\nu_\mu$ or $\bar \nu_\mu$ disappearance is observed, then the
NC $\pi^0$ and NC Elastic reactions can be used to determine whether the
disappearance is due to oscillations into active or 
sterile neutrinos. Oscillations into sterile neutrinos will result in a
suppression of events in the far detector, while oscillations into active
neutrinos will result in no suppression. Due to the
high statistics of the NC $\pi^0$ and NC Elastic event samples, the
statistical error will be small compared to the systematic error. The
sensitivity at 90\% CL for oscillations into sterile neutrinos at $\Delta m^2
\sim 1$ eV$^2$ is estimated to be 
$\sin^22\theta_{\mu\mu} \sim 3\%$ for neutrinos and 
$\sin^22\theta_{\mu\mu} \sim 5\%$ for antineutrinos.

\section{Other Experiments}

\begin{figure}
\centerline{\includegraphics[height=4.0in]{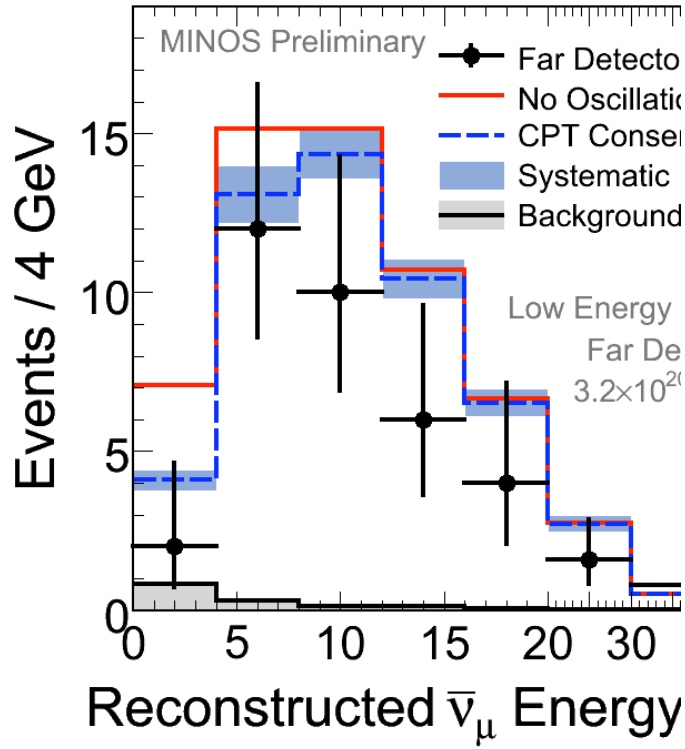}}
\caption{\label{numubar_minos} \em 
The  MINOS $\bar \nu_\mu$ event rate as a function of reconstructed
neutrino energy compared to the Monte Carlo expectation. Although the
statistics are low, the MINOS $\bar \nu_\mu$ event rate is suppressed
above 10 GeV, where there should be almost no suppression due to
atmospheric neutrino oscillations.}
\end{figure}

\begin{figure}
\centerline{\includegraphics[height=4.0in]{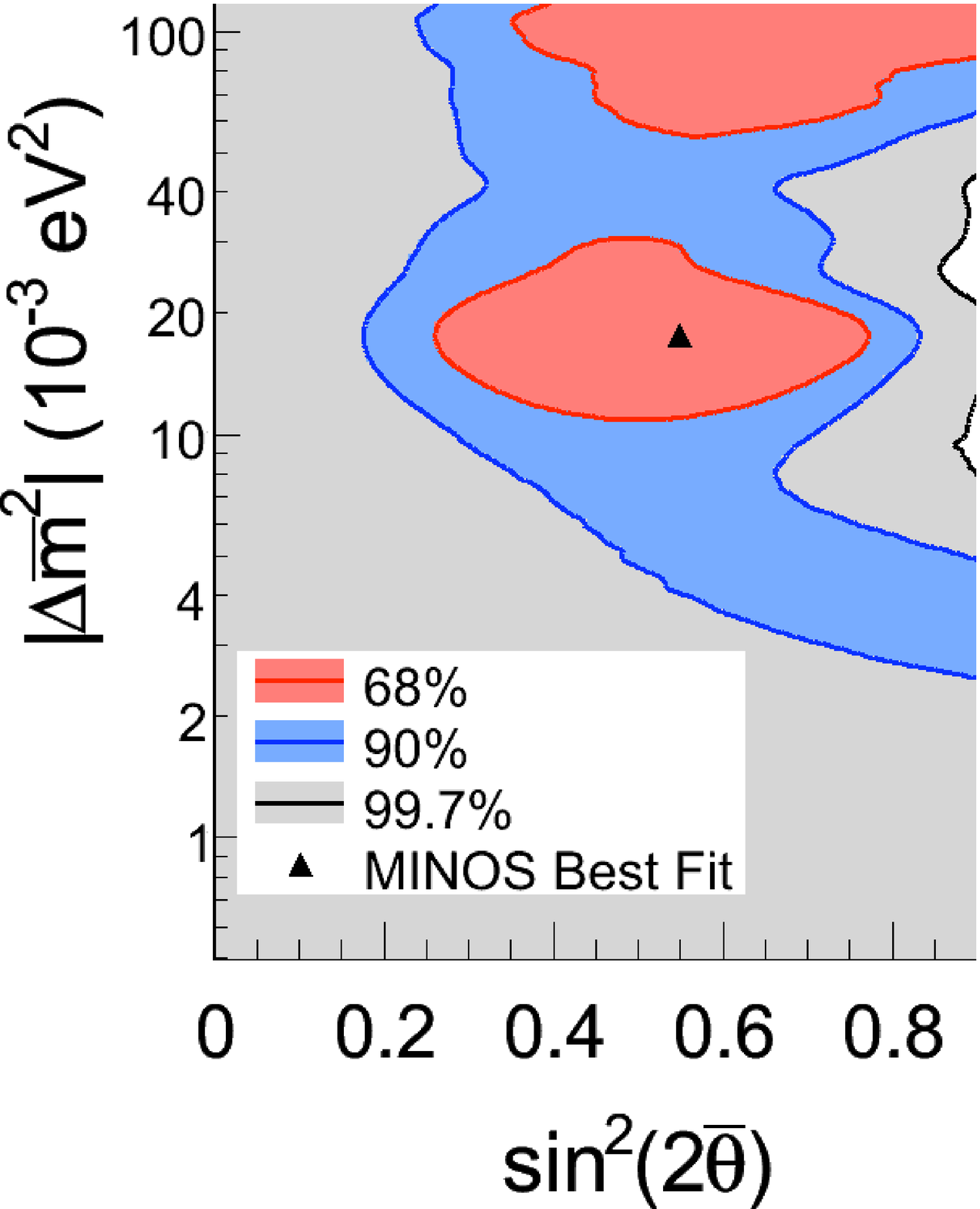}}
\caption{\label{numubar_allowed_minos} \em
The MINOS $\bar \nu_\mu$ disappearance allowed region, which is consistent
with oscillations at the $\sim 1$ eV$^2$ scale.}
\end{figure}

The MINOS experiment has measured neutrino oscillations at the
atmospheric scale with $\nu_\mu$ disappearance \cite{minos06} and has
begun to look at $\bar \nu_\mu$ disappearance \cite{minos09}. It is 
interesting to note that the MINOS antineutrino data so far are consistent with
the antineutrino 3+1 model of reference \cite{georgia} and consistent
with $\bar \nu_\mu$ disappearance at the LSND scale. Fig. \ref{numubar_minos}
shows the MINOS $\bar \nu_\mu$ event rate as a function of reconstructed
neutrino energy compared to the Monte Carlo expectation. Although the
statistics are low, the MINOS $\bar \nu_\mu$ event rate is suppressed
above 10 GeV, where there should be almost no suppression due to
atmospheric neutrino oscillations. Fig. \ref{numubar_allowed_minos} shows
the MINOS $\bar \nu_\mu$ disappearance allowed region, which is consistent
with oscillations at the $\sim 1$ eV$^2$ scale. 

The only other experiment or proposal that is capable of addressing these
physics objectives at the $\sim 1$ eV$^2$ scale 
is a Letter of Intent to refurbish the CERN PS
neutrino beam and build two liquid argon detectors \cite{rubbia}. 
However, the proposed BooNE experiment, with the existing BNB, 
should be able to obtain results prior to the CERN experiment.

\section{\bf Cost Estimate}

Table \ref{tab_hywel} shows a breakdown of the cost estimate for constructing a
second BooNE detector in a near location. The estimate
is based on the MiniBooNE construction costs. The
total estimated cost is \$7.3M, including
contingency ($\sim 30\%$) and escalation (3\% per year). The BooNE
construction is assumed to start in 2010 and last for 3 years.
The estimated cost for moving
MiniBooNE to a near location is $\sim \$4$M. An additional advantage of
moving MiniBooNE is that the MicroBooNE detector 
could then be moved into the original MiniBooNE enclosure, thereby
saving the expense of building a new MicroBooNE enclosure.

\begin{table}
%\vspace{5mm}
\centering
\begin{tabular}{|c|c|c|c|c|}
\hline
Item & Cost (\$K) \\
\hline
\hline
Tank and support structure     & 1065 \\ \hline
PMT's     & 1759 \\ \hline
Electronics/DAQ &  512   \\ \hline
Oil &     1429 \\ \hline
Calibrations & 412 \\ \hline
Miscellaneous &    198 \\ \hline
Engineering \& Construction & 1894 \\ \hline \hline
Total &  7269 \\ \hline
\end{tabular}
%\hspace{1in}
\caption{
A breakdown of the cost estimate for constructing a
second BooNE detector in a near location,
including contingency and escalation.
}
\label{tab_hywel}
\end{table}

\end{document}